\documentclass[a4paper,fleqn,usenatbib]{mnras}
\usepackage{newtxtext,newtxmath}

\usepackage[T1]{fontenc}
\usepackage{ae,aecompl}

\usepackage{graphicx}	
\usepackage{amsmath}	
\usepackage{amssymb}	
\usepackage{lscape}
\usepackage{color}

\newcommand{\kep}{\textit{Kepler} }

\newcommand{\jkt}{\textsc{jktebop} }

\newcommand{\vfit}{\textsc{v2fit} }

\newcommand{\ms}{m~s$^{-1}$}

\title[HIDES spectroscopy of \kep DEBs]{HIDES spectroscopy of bright detached eclipsing
binaries from the \kep field -- I. Single-lined objects.
}
\author[K. G. He{\l}miniak~et~al.]
{K. G. He{\l}miniak$^{1}$\thanks{Subaru research fellow, e-mail: xysiek@naoj.org},
N. Ukita$^{2,3}$,
E. Kambe$^{2}$,
S. K. Koz{\l}owski$^{4}$,
P. Sybilski$^{4}$,
\and
M. Ratajczak$^{5,4}$,
H. Maehara$^{2}$,
M. Konacki$^{4}$
\\
$^{1}$Subaru Telescope, National Astronomical Observatory of Japan, 650 North Aohoku Place, Hilo, HI 96720, USA\\
$^{2}$Okayama Astrophysical Observatory, National Astronomical Observatory of Japan, 3037-5 Honjo, Kamogata, Asakuchi,\\Okayama 719-0232, Japan\\
$^{3}$The Graduate University for Advanced Studies, 2-21-1 Osawa, Mitaka, Tokyo 181-8588, Japan\\
$^{4}$Nicolaus Copernicus Astronomical Center, Department of Astrophysics, ul. Rabia\'nska 8, 87-100 Toru\'n, Poland\\
$^{5}$Instytut Astronomiczny, Uniwersytet Wroc{\l}awski, ul. Kopernika 11, 51-622 Wroc{\l}aw, Poland\\
}

\date{Accepted XXX. Received YYY; in original form ZZZ}

\pubyear{2016}

\begin{document}
\label{firstpage}
\pagerange{\pageref{firstpage}--\pageref{lastpage}}
\maketitle

\begin{abstract}
We present results of our spectroscopic observations of nine detached eclipsing binaries (DEBs), selected from the \kep Eclipsing Binary Catalog, that only show one set of spectral lines. Radial velocities (RVs) were calculated from the high resolution spectra obtained with the HIDES instrument, attached to the 1.88-m telescope at the Okayama Astrophysical Observatory, and from the public APOGEE archive. 

In our sample we found five single-lined binaries, with one component dominating the spectrum. The orbital and light curve
solutions were found for four of them, and compared with isochrones, in order to estimate absolute physical parameters and evolutionary status of the components. For the fifth case we only update the orbital parameters, and estimate the properties of the unseen star. Two other systems show orbital motion with a period known from the eclipse timing variations (ETVs). For these we obtained parameters of outer orbits, by translating the ETVs to RVs of the centre of mass of the eclipsing binary, and combining with the RVs of the outer star. Of the two remaining ones, one is most likely a blend of a faint background DEB with a bright foreground star, which lines we see in the spectra, and the last case is possibly a quadruple bearing a sub-stellar mass object.

Where possible, we compare our results with literature, especially with results from asteroseismology. We also report possible detections of solar-like oscillations in our RVs.

\end{abstract}


\section{Introduction}
The launch of space photometric missions, such as CoRoT, MOST, or {\it Kepler}, 
has revolutionized many branches of stellar astrophysics. Extremely precise and
nearly continuous photometric observations led to many discoveries and improvements 
in studies of various kinds of variable stars. One of the fields that benefited 
the most are detached eclipsing binaries (DEBs). These are one of the most 
important objects for the astrophysics, as they allow to directly determine a 
number of fundamental stellar parameters, that are difficult or sometimes 
impossible to obtain otherwise. The unprecedented photometric quality of \kep 
data has opened new possibilities in studies of DEBs, as it allows for better than before 
precision in derivation of light-curve-based parameters, like fractional radii 
or orbital inclination, and observations of various other phenomena, for example: 
differential rotation, activity cycles, pulsations, including tidally induced, 
solar-like oscillations, etc. The \kep mission has revealed a zoo of objects, 
showing variety of characteristics.

The photometry alone, however, is not enough to accurately characterize a given
DEB. One has to combine it with information obtained from spectra, optimally taken
with high resolution. This information includes for example: precise radial 
velocities (RVs), monitored in time in order to observe their variability, or 
atmospheric parameters obtained from spectral analysis. Only such a complete
set of data is useful for the purposes of modern stellar astrophysics. 
Studies of DEBs are important not only because of the information one can obtain 
for a particular object. Characterisation of a large number of targets is 
beneficial for statistical studies, from which one can draw conclusions about,
e.g., the amount of binaries and higher-order multiples, or mechanisms
of their formation and evolution.

For this reason, we have started a spectroscopic monitoring of the brightest 
\kep DEBs with the 1.88-m telescope of the Okayama Astrophysical Observatory. 
For the purpose of precise characterization of stellar components,
one should target those DEBs that are also double-lined spectroscopic binaries 
(SB2s). However, it is also possible derive many important information 
from single-lined (SB1) objects. In this study we present a sample of 
such systems, showing the variety they represent -- from primary components 
of exotic binaries, to multiples of different orders. This work is organized as
follows. In Section \ref{sec_targets} we briefly present the target selection
criteria for our whole program, and basic information about the systems 
presented here. Section \ref{sec_data} presents the data used for this study.
Section \ref{sec_analysis} presents all the methods used for this work,
including spectra reduction, derivation of RVs and eclipse timing measurements, 
modelling RV and light curves, and comparison with isochrones. The results, are
presented in Section \ref{sec_results}, and the last Section contains a
short summary.

\section{Targets}\label{sec_targets}

\begin{table*}
\centering
\caption{KEBC information about the observed targets, and their character}\label{tab_cat}
\begin{tabular}{llllllcll}
\hline
\hline
KIC No. & Other name & RA (deg) & DEC (deg) & $P$~(d)$^a$ & $T_0^a$ (BJD-2450000) & $T_{eff}$ & $kmag$ & Char.$^b$ \\
\hline
03120320 & TYC~3134-38-1  & 292.2920 & 38.2845 & 10.2656134 & 4959.222315 & 5676 & 10.885 & SB1 \\
04758368 & TYC~3139-815-1 & 294.9364 & 39.8525 &  3.7499355 & 4958.206761 & 4594 & 10.805 & 3 \\
05598639 &WDS~J18551+4051AB&283.7774 & 40.8420 &  1.2975514 & 4955.013385 & N/A$^c$& 10.201 & 4? \\
08430105 & TYC~3146-491-1 & 291.5586 & 44.4882 & 63.3271056 & 4976.635546 & 4965 & 10.420 & SB1 \\
08718273 & TYC~3162-479-1 & 300.7676 & 44.8306 &  6.9590009 & 4955.951073 & 4577 & 10.565 & BL \\
10001167 & TYC~3546-941-1 & 286.9557 & 46.9366 &120.3909714 & 4957.586519 & 4683 & 10.050 & SB1 \\
10015516 & TYC~3560-2501-1& 293.4736 & 46.9704 & 67.6920522 & 5005.381429 & 5157 & 10.700 & SB1 \\
10614012 & TYC~3561-1138-1& 296.9287 & 47.8830 &132.1673120 & 4982.134994 & 4859 &  9.715 & SB1 \\
10991989 & TYC~3562-912-1 & 298.5271 & 48.4186 &  0.9744780 & 4954.647910 & 5021 & 10.282 & 3 \\
\hline
\end{tabular}\\
$^a$ For the eclipsing binary, where $T_0$ is the primary eclipse mid-time.\\
$^b$ ``SB1'': single-lined spectroscopic eclipsing binary;
``3'': tertiary companion to an eclipsing binary;\\ ``BL'': blend; ``4?'': possible quadruple.
$^c$ No temperature given in the catalogue.
\end{table*}

We have selected our targets from the \kep Eclipsing Binaries Catalog 
\citep[KEBC;][]{prs11,sla11,kir16}\footnote{http://keplerebs.villanova.edu/}, 
by putting the following criteria:
\begin{enumerate}
\item \kep magnitude $kmag < 11$, to have the targets within the brightness range of 
the telescope.
\item Morphology parameter \citep{mat12} $morph < 0.6$ to exclude 
contact and semi-detached configurations.
\item Effective temperature $T_{eff} < 6700$~K, as given in KEBC, to have 
only late type systems, with many spectral features. When the $T_{eff}$ was not 
given, we were checking the available colour or spectral type information, 
and selecting stars redder than $V-K=1.0$ and of spectral type later than~F0.
\end{enumerate}
We have pre-selected 75 targets, 20 of which have been observed so far. One has
been already described in \citet{hel15b}, and nine are 
presented in this work. These are systems whose spectra show
only one set of lines. In general, they can be divided in two groups: (i) when
the motion does not coincide with eclipses, which means that the total flux is 
dominated by a third source, either gravitationally bound with the EB (a circumbinary
companion), or not (a blend); and (ii) when the observed motion coincides with the 
period of the eclipses, meaning that one of the eclipsing components is dominating 
the total light of the system. Systems from group (i) will be hereafter called
``third stars'' (3rd), or ``blends'' (BL), and those from group (ii) will be called.
``single-lined binaries'' (SB1). The systems with both eclipsing components seen
in the spectra (SB2), whose orbital motion may be traced and a full set of physical 
parameters found from a joint light and radial velocity curves analysis, will be 
presented in another paper.

The targets presented in this study are summarized in Table \ref{tab_cat}. 
For each of them we briefly present the basic information below.
The eclipsing nature of all targets was discovered by the \kep mission.
Unless stated otherwise, no radial velocity data have been published till date. 
The targets, obviously, appear in several catalogue papers related to the
\kep mission \citep[like][]{cou11,ten12}, but information from these are omitted,
due to unknown influence of the light from other component(s) in the system
on the results of the analysis.

\begin{description}
\item {\it KIC~03120320 = TYC~3134-38-1}:
This target has the highest KEBC effective temperature and lowest brightness 
in the sample. Narrow and highly uneven eclipses (14.3 and 1.3 per cent), 
suggest domination of one of the components over the other. Except brightness and 
position measurements, no literature data is available.

\item {\it KIC~04758368 = TYC~3139-815-1}:
This target was recognized by \citet{con14} as showing eclipse timing variations,
(ETV) with a long, poorly-constrained period. Shallow eclipses ($\sim$3.5~per~cent) 
suggest that the eclipsing pair is dominated by a bright, third star. Indeed, RV
measurements seem to coincide with the ETV period, rather than with the eclipses. 
The high brightness, low effective temperature and gravity ($\log(g)=2.624$) implicate 
that the outer star is a red giant. It is, however, listed in \citet{gau13} as not
showing solar-like oscillations.

\item {\it KIC~05598639 = KOI~6602, WDS~J18551+4051AB, HEI~73}:
Classified in the \kep Object of Interest (KOI) 
catalogue\footnote{http://exoplanetarchive.ipac.caltech.edu/cgi-bin/TblView/nph-tblView?app=ExoTbls\&config=koi}
as a planetary candidate (PC; event KOI~6601.01).
It is a visual pair discovered by \citet{hei80}. The Washington Double Star 
Catalog \citep[WDS;][]{mas01} contains 6 position measurements, taken
between 1979 and 2008, covering only a very small fraction of the orbit. 
The last measurement places the fainter component 0.7 arcsec away from the
brighter, at position angle of 319 degrees. The WDS also gives magnitudes of components: 
10.95 and 11.10 mag. Influence of the third light causes the nearly-equal eclipses to be 
relatively shallow ($\sim$11 per cent). Before our observations it was not certain which 
of the two components is the eclipsing pair. Presence of only one set of lines in the 
spectrum is also puzzling. \citet{con14} calculated ETVs for this system, but did not 
draw any conclusions from them. It is also listed as observed with the SOPHIE
spectrograph by \citet{san16}, but no RV measurements are given, nor are spectra 
available in the archive.

\item {\it KIC~08430105 = KOI~3873, TYC~3146-491-1}:
Classified in the KOI catalogue
as a false positive (FP: significant secondary flag; event KOI~3873.01). 
This system shows a complete primary (deeper)
eclipse, suggesting that the colder star is significantly larger than the hotter component,
and dominates the spectrum. It was recognized by \citet{gau13,gau14} as a red giant
that shows solar-type oscillations. From asteroseismology, \citet{gau14} estimated 
its mass and radius to be 1.04(12)~M$_\odot$ and 7.14(28)~R$_\odot$, respectively.

\item {\it KIC~08718273 = KOI~5564, TYC~3162-479-1}:
Classified in the KOI as a planetary candidate (PC; event KOI~5564.01). 
This system shows slightly unequal eclipses, the shallowest in our sample 
($\sim0.3$~per~cent), suggesting a dominant third light. Spectra show only 
one set of lines, relatively stable in velocity over the course of the observations. 
This star is a 1.9(3)~M$_\odot$, 15.2(8)~R$_\odot$ red giant that shows solar-type 
oscillations \citep{gau13}.

\item {\it KIC~10001167 = TYC~3546-941-1}:
Another system with complete primary eclipse, hence the colder, much larger component 
dominates over the warmer one. \citet{gon08} suggested that it is a red clump giant,
and estimated the distance to be 705~pc. Also recognized by \citet{gau13,gau14} as 
showing solar-type oscillations. From asteroseismology, \citet{gau14} estimated 
its mass and radius to be 1.13(07)~M$_\odot$ and 13.85(32)~R$_\odot$, respectively.

\item {\it KIC~10015516 = KOI~990, TYC~3560-2501-1}: 
Classified in the KOI catalogue as a false positive (FP; event KOI~990.01).
This target also shows a complete primary eclipse.
Except brightness and position measurements, no literature data is available.

\item {\it KIC~10614012 = TYC 3561-1138-1}:
This system has the longest period of eclipses in our sample. It is also the brightest
one. The star was recognized as a pulsating ``heartbeat'' red giant by \citet{bec14},
who give 1.49(8)~M$_\odot$, 8.6(2)~R$_\odot$. Orbital solution based on the 
observations with the HERMES spectrograph is also given, but without any data points.

\item {\it KIC~10991989 = TYC 3562-912-1}:
This system has the shortest eclipsing period in our sample, and was also recognized
as showing eclipse timing variations \citep{rap13,con14} with the period of 554.2~d.
Very shallow eclipses (primary: 0.8, secondary 0.5 per cent) suggest domination of 
the third star in the total flux. Only one set of lines is identified, and its motion 
seems to coincide with the ETV period. The star visible in the spectra
was identified as a pulsating red giant by \citet{gau13}, who estimated its mass
and radius: 2.5(4)~M$_\odot$, 9.6(5)~R$_\odot$.
\end{description}

\section{Data}\label{sec_data}

\subsection{HIDES Observations}
The spectroscopic observations were carried out at the 1.88-m telescope of the 
Okayama Astrophysical Observatory (OAO)
with the HIgh-Dispersion Echelle Spectrograph \citep[HIDES;][]{izu99}. The instrument 
was fed through a circular fibre, for which the light is collected via a circular 
aperture of projected on-sky diameter of 2.7 seconds of arc, drilled in a flat mirror 
that is used for guiding \citep{kam13}. An image slicer is used in order to reach 
both high resolution ($R\sim50000$) and good efficiency of the system. 

The observations were made between July 2014 and June 2016, under various weather
conditions. The wavelengths calibration was done on the basis of ThAr lamp exposures
taken every 1-2 hours. With this set-up and observing strategy we reached the RV 
precision of 40 -- 60~\ms, as measured from observations of four velocity standards 
(note similar level of precision in some of our orbital solutions). The December 2014 run 
suffered from a change in the spectrograph's set-up, which resulted in different echelle 
format, wavelength solution, and velocity zero-point. Fortunately, this has been accounted for 
by observations of two of the standards -- all the velocities from 
that time were corrected, and the reported stability already includes this run.
Most of the 2016 observations were done under a shared-risk mode, using a new
queue scheduling software of the OAO-1.88 telescope.

The exposure times varied from 600-900~s for the brightest to 1500-1800~s 
for the faintest stars (2400~s once). The signal-to-noise ratio
varied usually between 30 and 60.

\subsection{Publicly available data}
Other data used in this study is publicly available. 
The long cadence \kep photometry for all targets 
is available for download from the KEBC. We used the de-trended relative flux measurements $f_{dtr}$, 
that were later transformed into magnitude difference $\Delta m=-2.5 \log(f_{dtr})$,
and finally the catalogue value of $k_{mag}$ was added. 

Five of our targets -- KIC~03120320, 04758369, 08430105, 08718273, and 10001167 
-- were observed by the APOGEE survey \citep{all08,maj15}. 
We have extracted RV measurements from the visit 
spectra\footnote{http://dr12.sdss3.org/advancedIRSearch}, 
and used them together with our data. They are listed in Table 
\ref{tab_RV_apo}, together with their unique identifiers that are 
composed of Plate IDs, MJD and fibre numbers. 

\section{Analysis}\label{sec_analysis}

\subsection{CCD reduction and spectra extraction}
The HIDES instrument records data on three chips, all being mosaics
of two 1024$\times$4100~pix detectors, with extra 50+25 columns and 15 rows 
being read out of overscan.
The reduction was made using dedicated \textsc{iraf}-based scripts that deal with 
all chips simultaneously. The basic reduction steps include correction 
for bias and ``camera'' flat field. For this step we used additional exposures 
of a halogen lamp, taken in the slit mode, using slits of two lengths: 1.4 (for 
two ``redder'' chips) and 1.25 arcsec. We first combined them into a master 
flat, then made a ``smooth'' flat, using a running boxcar with a 100 pix window along 
the dispersion direction. Finally the ``master'' is divided by the ``smooth'', 
leaving only the inhomogeneities from the camera optics in the areas where spectra 
are recorded. Data were also corrected for bad pixels, cosmic rays and scattered light, 
and then spectra were extracted. Due to crowding of the 
apertures and sudden drop in the signal, from the bluest chip we were extracting 
only the 15 reddest orders. After assigning a wavelength solution to each chip
separately, three spectra from different chips were then merged into a single
echelle exposure. The final product is composed of 53 spectral orders, 
spanning from 4360 to 7535 \AA, i.e. not covering the calcium H and K lines.

\subsection{HIDES RV measurements}
For the radial velocity measurements we used our own implementation of the 
classic cross-correlation function (CCF) technique, that compares the observed
spectrum with a synthetic template in the velocity domain.
As templates we used synthetic spectra computed with ATLAS9 and ATLAS12 codes 
\citep{kur92}, which do not reach wavelengths longer than 6500~\AA, thus 
only 30 echelle orders (4360--6440~\AA) were used for RV calculations.
We calculated CCFs for each order separately, and later merged them into one.
Single measurement errors were calculated with a bootstrap approach, in a
similar way as in \citet{hel12}. In this and other studies we found that 
these errors are sensitive to the signal-to-noise ratio ($S/N$) of the spectra 
and rotational broadening of the lines, thus are very good for weighting the 
measurements during the orbital fit.

All radial velocity measurements obtained from our HIDES spectra, together
with their errors and $S/N$ of the spectra, are
listed in Table \ref{tab_RV} in the Appendix.

\subsection{Eclipse timing variations}\label{sec_etv}
Thanks to the unprecedented quality of \kep photometry, it is possible
to obtain very precise moments of minima of eclipsing binaries, and search
for periodicities in eclipse timing variations (ETVs). Several such studies have 
been already made \citep{bor15,bor16,con14,gie12,gie15,rap13}, and
three of our systems have their ETVs reported.
These are KIC~04758368 \citep{con14,bor16}, 05598639 \citep{con14}, and 10991989 
\citep{rap13,con14,bor16}. Only \citeauthor{con14} make their measurements available.
A quick comparison of their data and figures provided by other authors reveals that 
for KIC~04758368, for example, the O-C graphs look different. Also, due to a very long 
outer period, the orbital solutions are highly uncertain in this case. The ETVs for 
05598639 are showing some strong, uncorrected systematic effects, which is the most 
likely reason why no ETV-based orbital solution for this system has been presented.
Finally, the available solutions for 10991989 are essentially different, 
e.g. \citet{rap13} and \citet{bor16} find a substantial eccentricity, while 
\citet{con14} assume a circular orbit.

To be able to analyse these objects in a homogeneous way, we decided to calculate 
our own ETVs. We used the radio-pulsar-style approach, presented in \citet{koz11}.
In this method, a template light curve is created by fitting a trigonometric
(harmonic) series to a complete set of photometric data. Then, the whole set of 
photometric data is divided to a number of subsets (the number is arbitrary). 
For each subset the phase/time shift is found by fitting the template curve with a 
least-squares method. This approach is well suited for large photometric data sets, 
especially those obtained in a regular cadence, like from the \kep satellite.

\subsection{Translating ETVs to RVs}\label{sec_oc2rv}
The information possible to obtain from the eclipse timing variations
have a similar character as from the radial velocities, i.e. it is the
radial component of the orbital motion that is observed. It is thus possible
to translate the ETVs of the inner eclipsing binary to the RVs of its 
centre of mass (variations of its systemic velocity). In this section
we will only refer to the ``outer'' orbit in a hierarchical triple system,
which is the orbit of the tertiary star and the inner eclipsing binary
around their common centre of mass (CCM). 

The length of the orbital position vector $r$ is given by:
\begin{equation}
r = a \frac{1-e^2}{1+e \cos{\nu}},
\end{equation}
where $a$ is the major semi-axis, $e$ is the orbital eccentricity, 
and $\nu$ is the true anomaly. The radial (in the direction of the
line of sight) component $z$ is:
\begin{equation}
z = r \sin{(\omega+\nu)} \sin{i},
\end{equation}
where $\omega$ is the longitude of the pericentre and $i$ the orbital
inclination. During the orbital motion, when the EB is closer to
or further from the observer, the eclipses appear earlier or later 
by the amount of time $\tau_{lte}$ the light (travelling with the speed $c$) needs 
to take to travel the distance $z$:
\begin{equation}\label{eq_roe}
\tau_{lte} = \frac{z}{c} = \frac{a\sin{i}}{c} \frac{(1-e^2)\sin{(\omega+\nu)}}{1+e \cos{\nu}}.
\end{equation}
This is the formula for the classical, light travel time or R\"{o}mer delay, 
where the amplitude of the effect is 
\begin{equation}\label{eq_a}
A = \frac{a\sin{i}}{c}.
\end{equation}
At the same time the, radial velocity $v_r$, by definition, is the time
derivative of the radial component of the distance:
\begin{equation}
v_r = \dot{z} = c \dot{\tau}_{lte}.
\end{equation}
Calculation of the time derivative of $\tau_{lte}$ from the 
Equation~\ref{eq_roe} leads to the classical formula for the radial velocity:
\begin{equation}
v_r = \frac{2 \pi a \sin{i}}{P \sqrt{1-e^2}} (e\cos{\omega} + \cos{(\omega+\nu)}),
\end{equation}
where $P$ is the orbital period, and the RV amplitude $K$ is given by:
\begin{equation}\label{eq_k}
K=\frac{2 \pi a \sin{i}}{P \sqrt{1-e^2}},
\end{equation}
and where only the systemic velocity $v_\gamma$ is omitted. 
Combination of Equations \ref{eq_a} and \ref{eq_k} 
gives the formula that joints the amplitudes of the two phenomena:
\begin{equation}\label{eq_ak}
\frac{K}{A} = \frac{2\pi c}{P\sqrt{1-e^2}}.
\end{equation}
It may be directly used when the orbital solution from ETVs is
already known, in order to predict the scale of the RV variations.
However, from the three systems in our sample that show ETVs, 
KIC~10991989 does not have it modelled, and solutions
for KIC~04758368 are different in \citet{rap13} and \citet{con14},
thus we have developed a general approach that works on crude
ETV measurements and takes into account their spread, and does not 
assume or require orbital elements to work.

\begin{figure*}
\includegraphics[width=0.95\textwidth]{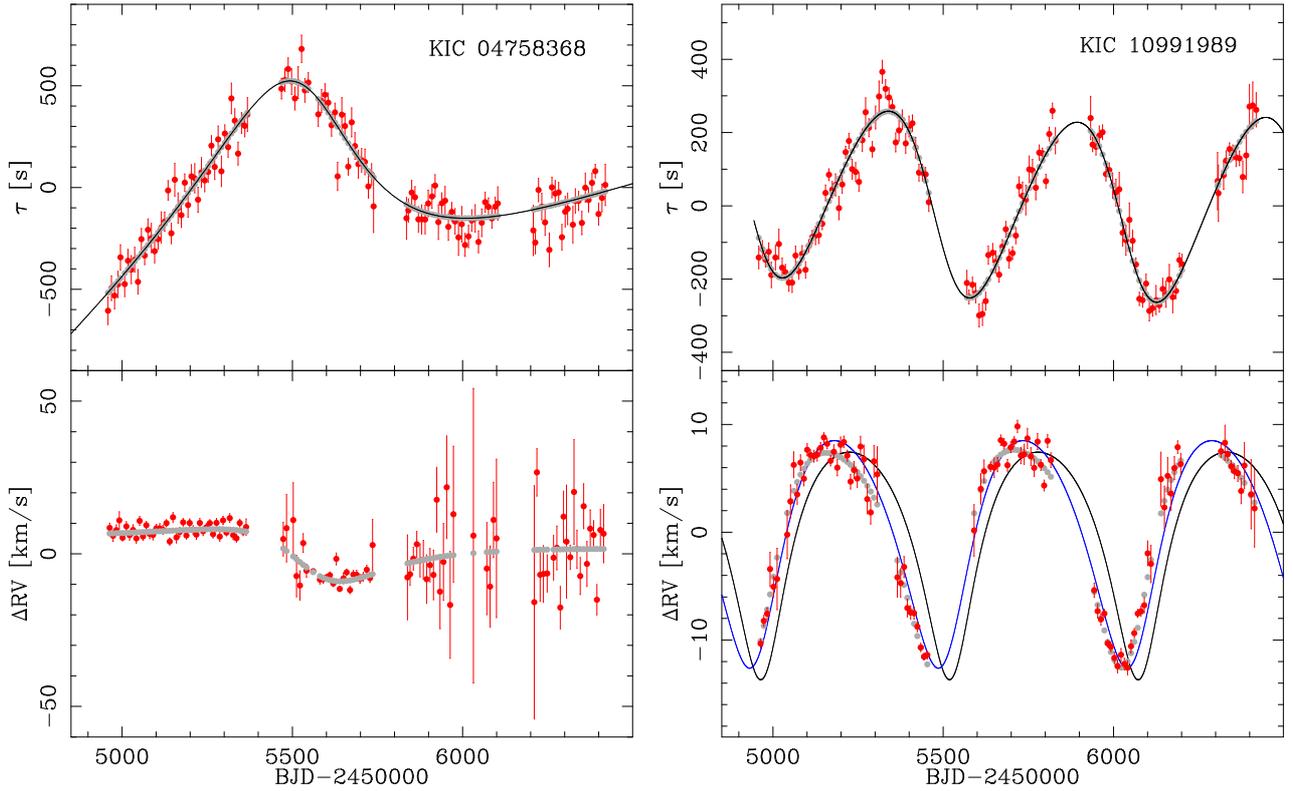}
\caption{Illustration of the method of translating ETV measurements
to RVs, on the ETV data for KIC~04758368 (left) and KIC10991989 (right). 
{\it Top}: Red points show our ETV measurements $\tau^i$, 
the black line is the generic fit to these data, and the grey dots are the model 
ETVs $\tau_0^i$. {\it Bottom:} Reproduced RVs of the centre of mass of the eclipsing 
binaries in these systems with scatter (red) and without (model, grey points). For 
KIC~10991989 solid lines present theoretical RV curves for a set of orbital parameters 
representing the best-fitting solution from \citet{rap13} (black), and for a set with
values within their uncertainties (blue). The latter is in agreement with our 
reconstruction.}\label{fig_oc2rv}
\end{figure*}

Calculating numerical derivative of a noisy sequence of data,
like \kep's ETVs, is quite challenging. Thus first we take our
ETV measurements $\tau^i$ and fit a
generic model ETV curve, that has no physical meaning, and is used only to
get the shape of the timing variations and their probable ``model''
values $\tau_0^i$. This function can be periodic, but can also be a simple 
polynomial or spline (useful when ETVs do not cover the full orbital 
period). Having the ``model'' ETVs tabulated, it is fairly easy to calculate 
their derivatives and obtain the ``model'' radial velocity measurements.
We do this with the simple two-point approach, for two consecutive 
ETVs $\tau_0^i$, $\tau_0^{i+1}$, measured in times $t^i$ and $t^{i+1}$: 
\begin{equation}
v_{r,0}^i = c\frac{\tau_0^{i+1} - \tau_0^i}{t^{i+1} - t^i},
\end{equation}
except for breaks between observations longer than 10 days. The time stamp
is the mid-time of the two measurements, and $c$ is the speed of light. 
Now, we introduce the scatter to each of the model RVs. Intuitively, it
is dependent on the difference between two measured ETVs. We define
the correction to the $i$-th ``model'' RV $\epsilon_v^i$ = $v^i-v_0^i$
as the fraction of the measured and model ETV differences, multiplied by
$c$ and the time-related term (as in the previous Equation):
\begin{equation}
\epsilon_v^i = c\frac{\tau^{i+1}-\tau^i}{\tau_0^{i+1} - \tau_0^i} \frac{1}{t^{i+1} - t^i}.
\end{equation}
In this way we make the scatter of the RVs dependent on the scatter
of the ETVs.
We then tabulate the time stamps $(t^{i+1} + t^i)/2$, scattered RVs
$v_r^i = v_{r,0}^i + \epsilon_v^i$. We also calculate RV uncertainties $\sigma_v^i$,
because in general each ETV can be given with a different error $\sigma_\tau^i$:
\begin{equation}
\sigma_v^i = c\frac{\sqrt{(\sigma_\tau^{i+1})^2 + (\sigma_\tau^i)^2}}{|\tau_0^{i+1} - \tau_0^i|} \frac{1}{t^{i+1} - t^i}.
\end{equation}
Finally, we run two iterations of a sigma-clipping procedure to remove 
the obvious outliers. Their number does not exceed 10 per cent of all post-ETV
radial velocities. The resulting RVs are given in Table \ref{tab_RV_etv},
and our original ETVs in Table \ref{tab_ETV}, both in the Appendix.

Our approach and its result is seen in the Figure~\ref{fig_oc2rv}.
It works very well for periodic cases, like KIC~10991989, where ETVs cover
several periods of the outer orbit. We can compare our reconstructed RVs 
to a model based on the solution from \citet{rap13}, who give
$P=554.2$~d, $e=0.30^{+0.07}_{-0.09}$, $A=256^{+18}_{-17}$~s, which leads to 
$K=10.56\pm0.61$ k\ms (ratio of two amplitudes: $K/A = 0.041(4)$). 
One can see that our method gives RVs in agreement with this solution --
in Fig.~\ref{fig_oc2rv} the blue line represents a solution with parameters
within the uncertainties given by \citeauthor{rap13}, and it 
passes through our post-ETV velocities. Analogous comparison can 
not be made for KIC~04758368, for which the solutions are highly
uncertain due to a very long period, nor for KIC~05598639, for which
no attempt to solve the ETVs have been made.

One can see that many post-ETV velocities of KIC~04758368 have large error and
show a scatter comparable to or even larger than the RV amplitude itself 
(as seen from the model RVs), especially for BJD$>$2455800. This is actually
expected when the scatter of ETVs is comparable to or larger than the
scale of their variation. Radial velocities are time derivatives of the ETVs, so the more
stable ETVs the less precise RVs. Our method also has the potential
to verify the ephemeris of the inner eclipsing binary. Incorrect $P_{ecl}$
will lead to a linear trend in ETVs that may not be easily seen in 
some cases. In such case the post-ETV velocities will show a constant shift
(non-zero $\gamma$), that may be easily detectable. A change in $P_{ecl}$ 
(meaning a non-zero $\dot{P}_{ecl}$), that may originate from
a mass transfer for example, will produce a linear trend in RVs.

Note that our approach includes only the R\"omer time delay,
not the longer time-scale perturbations, like the physical delay
\citep{rap13,bor15}. Similar calculations can of course be done for timing of 
other periodic phenomena, like pulsations.

\subsection{Orbital solutions}
The orbital solutions were found with our own procedure called {\sc v2fit}
\citep{kon10}. This code fits a single or double-keplerian orbit to a set
of RV measurements of one or two components, utilizing the 
Levenberg-Marquard minimization scheme. The simplest version fits
the period $P$, zero-phase\footnote{Defined as the moment 
of passing the pericentre for eccentric orbits or quadrature for
circular.} $T_0$, systemic velocity $\gamma$, velocity semi-amplitudes $K_{1,2}$,
eccentricity $e$ and periastron longitude $\omega$, although in
the final runs the last two parameters were sometimes kept fixed on values
found by \jkt fit (see next \ref{sec_jkt}). The {\sc v2fit} code allows 
for some perturbations and secondary effects, such us 
differences between zero-points of multiple spectrographs, or 
difference between systemic velocities of two components, $\Delta\gamma \equiv \gamma_2-\gamma_1$,
which in this work was used for post-ETV RVs. The {\sc v2fit} can also fit 
for linear and quadratic trends in $\gamma$, and periodic modulations of
$\gamma$, interpreted as influence of a circumbinary body on an outer orbit.
In the most sophisticated version it also employs the Wilson-Devinney (WD) 
code to account for tidal distortions of both components, as well as a number 
of relativistic effects, but these corrections were not used 
in this work, nor were the perturbations of $\gamma$.
The code further calculates the projected values of major semi-axes 
$a_{1,2}\sin{i}$, and minimum masses $M_{1,2}\sin^3{i}$ for the SB2s.
Whenever applicable, we simplified our fit by keeping the orbital
period on the value given in the KEBC. Also, we were first 
letting $e$ and $\omega$ to be fit for, but when the resulting value of $e$
was indifferent from zero, we were keeping it fixed. Other parameters (when 
applicable) were always set free, unless clearly stated in the further text.

Formal parameter errors of the fit are estimated by forcing the final 
reduced $\chi^2$ to be close to 1, either by adding in quadrature 
a systematic term to the errors of RV measurements (jitter),
or multiplying them by a certain factor. Because the code weights the 
measurements on the basis of their own errors, which are sensitive to 
$S/N$ and $v_{rot}$, we used the second option in our analysis. The errors 
given in Tables \ref{tab_RV_apo}, \ref{tab_RV}, and \ref{tab_RV_etv} are the multiplied ones, 
for which $\chi^2 \simeq 1$. Systematics coming from fixing a certain parameter 
in the fit are assessed by a Monte-Carlo procedure, and other possible 
systematics (like coming from poor sampling, low number of measurements, 
pulsations, etc.) by a bootstrap analysis. All the uncertainties of orbital 
parameters given in this work already include the systematics.

\subsection{Light curve solutions}\label{sec_jkt}
To deal with large amounts of photometric data from the \kep 
telescope for multiple targets, one either needs a lot of
computer time, or a fast, efficient, and somewhat simplified 
code. For the purpose of this work we choose the second option, 
and made sure that the simplifications either do not affect
the final results or are properly taken into account in the
error budget. 

For the light curve (LC) analysis we used version 28 (v28) 
of the code \jkt \citep{sou04a,sou04b}, which is based on the 
\textsc{ebop} program \citep{pop81}. It is a fast procedure 
that works on one set of photometric data at a time, and in the
version we used, not allowing for analysis of RV curves. 
It also does not allow for spots or pulsations. It however offers
a number of algorithms to properly include systematics in the error
budget. The exquisite precision of \kep photometry reveals all
sort of out-of-eclipse brightness modulations in the DEBs, 
mainly spots and pulsations. In our objects, however, after years
of nearly continuous observations they are averaged out over the
orbital period. This results in a spread of data points
in the phase-folded LCs that is larger than the telescope's 
photometric precision (sometimes quite significantly). For this
reason we decided to treat them as a correlated noise and run the 
residual-shifts (RS) procedure to calculate reliable uncertainties 
\citep{sou11}. However, the time required to run the code in
RS mode scales as the square of the number of data points.

Therefore, for finding the best fit we used the whole curve, but for 
the errors estimation we decided to analyse data from each quarter
separately. {\it Kepler}'s long-cadence complete Q0-Q17 curves contain 
over 60,000 points, while single-quarter ones do not exceed 5000, so 
the time required to run the RS stage was at least 8 times shorter.
For the systems with longest periods ($>$100 days) we
sometimes merged two quarters into one set, in order to have both 
primary and secondary eclipse completely covered. 
We then calculated weighted averages of the resulting parameters.
To get final parameter errors, we added in quadrature the
formal error of the weighted average and the {\it rms} of the results
from each quarter or set. When the additional variability was of a time
scale much shorter than the length of a quarter (like oscillations), 
it made the weighted average error dominant. If the 
variability was of a longer time scale (like stable spots), it 
produced trends in parameters visible from quarter to quarter, and made 
the {\it rms} dominant. We believe that in such way we took care
of the systematics in the LC analysis.

We fitted for the period $P$, mid-time of the primary (deeper) minimum 
$T_{ecl}$, sum of the fractional radii $r_1 + r_2$ (where $r = R/a$), 
their ratio $k$, inclination $i$, surface brightness ratio $J$,
maximum brightness $S$, and third light contribution $L_3$ for multiples. 
In case of eccentric SB1 systems, which can be recognized by the secondary
eclipse located at a phase different than 0.5, we also fitted
for $e$ and $\omega$, as the low number of spectroscopic data does
not allow to constrain them securely. Their final values are from
the \jkt runs. The gravity darkening coefficients were always kept 
fixed at the values appropriate for stars with convective envelopes ($g = 0.32$).
Due to the nature of studied objects, we did not use any spectroscopic ratios,
which somewhat increased the resulting uncertainties in $k$.
At the end, the code calculates the fractional radii $r_{1,2}$, and 
fluxes $L_{1,2}$. 

\subsection{Absolute parameters and age from isochrones}
In case of those SB1 systems for which we performed the \jkt analysis,
we can estimate the absolute stellar parameters and age of a system
by comparing our partial results with theoretical isochrones.
For this purpose we chose the PARSEC set \citep{bre12}, which includes
calculation of absolute magnitudes in the \kep photometric band. 
We use the parameters from \jkt runs, i.e. $F = L_2/L_1$, $rr = r_1+r_2$, 
and $k = r_2/r_1$, and we look for such values of masses of the two stars 
and their age, that minimize the following term:
\begin{equation}
\chi^2 = \left( \frac{k_{is} - k_{JK}}{\sigma_k} \right)^2 +
\left( \frac{rr_{is} - rr_{JK}}{\sigma_{rr}} \right)^2 +
\left( \frac{F_{is} - F_{JK}}{\sigma_F} \right)^2,
\end{equation}
where index $is$ denotes the values for a given pair of stars found in the 
isochrones, and parameters with the index $JK$ and the corresponding 
uncertainties $\sigma$ are taken from the \jkt solutions. 
For a set of ages, the PARSEC isochrones list masses, gravities, 
effective temperatures, and absolute magnitudes in the \kep and SDSS bands. 
To calculate the fractional radii of the given pair of putative stars (to be
compared with \jkt results), we first calculated the major semi-axis from our 
value of period and the putative masses, and then the radii using the masses and 
gravities. The \kep band flux ratio $F = L_2/L_1$ was estimated from the 
difference of two putative absolute magnitudes. The search was performed for 
ages $t$ between $10^{6.6}$ and $10^{10.1}$~Gyr, with the step of 0.05 in 
$\log(t)$, with the additional condition that the secondary must be smaller, 
and either hotter or cooler than the primary, depending on the results of orbital 
fits. Sometimes we added an additional constrain regarding the SDSS $g'-r'$ colour.
Finally, from the difference of observed and absolute magnitudes $kmag-Kmag$ for 
the whole system, and the primary alone if a total eclipse is observed, 
we can estimate the distance to the pair.

The analysis described here was performed only for one value of metallicity,
which was taken from the MAST, and without any interpolation between nodes within 
a single isochrone, nor between isochrones. For this reason we did not attempt 
to estimate the uncertainties, and we treat the results as approximate, however 
self-consistent. The cited ages should be treated with caution, but the
evolutionary stages should be determined securely.

\section{Results}\label{sec_results}

\subsection{RV standards and stability}\label{sec_rvstd}

\begin{figure}
\includegraphics[width=\columnwidth]{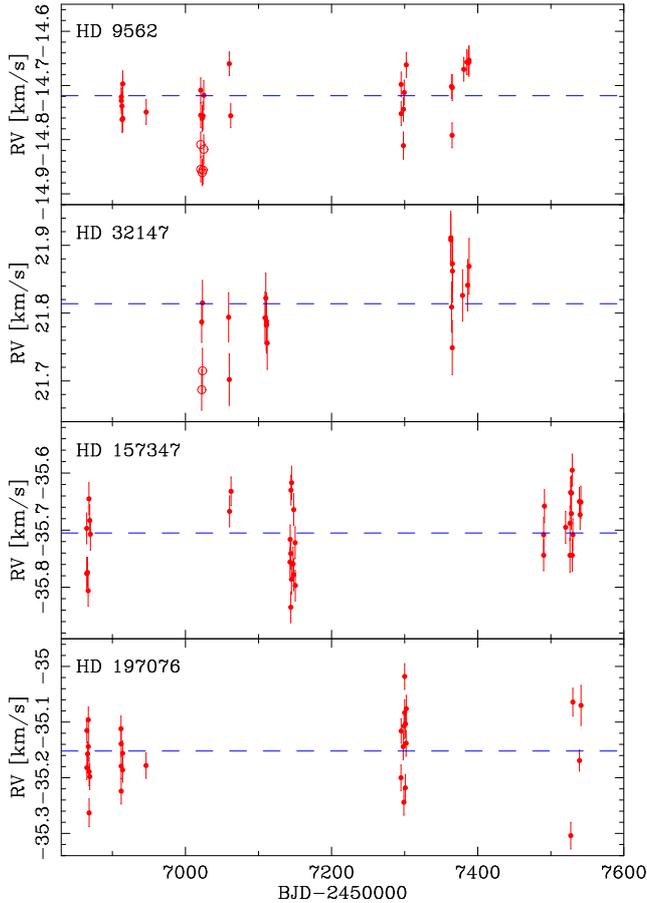}
\caption{Results of RV measurements of the standards stars.
Points taken for the analysis are marked with filled circles, 
while open symbols show values from December 2014 before the correction
(for HD~9562 and HD~32147 only).
Dashed lines mark the average values.}\label{fig_std}
\end{figure}

\begin{table}
\centering
\caption{Results for the radial velocity standards.
For each star its spectral type, $V$-band magnitude, average
RV ($\bar{v}$, in k\ms), $rms$ (in k\ms), number of measurements ($N$) 
and time span ($\Delta t$, in days) is given.}\label{tab_std}
\begin{tabular}{lccrrccr}
\hline \hline
Star & Sp.T. & $V$ & \multicolumn{1}{c}{$\bar{v}$} & $rms$ & $N$ & $\Delta t$\\
\hline
HD~9562   	& G1 V & 5.76 & -14.719 & 0.044 & 28 & 476 \\
HD~32147  	& K3 V & 6.21 &  21.814 & 0.054 & 19 & 366 \\
HD~157347	& G3 V & 6.29 &	-35.705 & 0.060 & 36 & 676 \\
HD~197076	& G5 V & 6.44 & -35.152 & 0.064 & 30 & 676 \\
\hline
\end{tabular}
\end{table}

The long-term stability of the spectrograph was monitored with observations
and RV measurements of four radial velocity standards: HD~9562, HD~32147, 
HD~157347, and HD~197076. The first two were observed during the unfortunate 
run when the instrument set-up was changed, and allowed for the spectrograph's
zero-point correction. For this purpose for each of the two stars we calculated 
the average velocities during the unaffected observations, and shifted all 
measurements by that value. We then combined the two sets and for both stars 
calculated the shift of the December 2014 measurements only: $-0.100(10)$~k\ms, 
and corrected them by this value. Finally, for both stars, we have calculated 
the average RVs and $rms$ again. Values of all velocities from the December run 
were then increased by 100~\ms.

The results of observations of standards are summarized in Table \ref{tab_std} 
and Figure~\ref{fig_std}. In case of the December 2014 run both corrected and 
uncorrected RVs are shown in the figure, but the final results from the table 
include only the corrected. We found the spectrograph to be stable to 
$\sim$40-60~\ms\,over the course of our observations, and even down to 30 \ms\,for 
shorter periods of 3-5 months. Some of the orbital solutions we found reach even 
lower $rms$, even below 20. Note that we were not using the iodine cell and not 
taking ThAr exposures before and after each science spectrum. For comparison, 
the precision achieved with HIDES with iodine cell is 5-20 \ms\,\citep{sat13a,sat13b}.

\subsection{No orbital RV motion}

In this section we present two objects where no orbital motion
has been detected in the RVs. These are KIC~08718273 -- a blend of a DEB with 
a foreground star, and KIC~05598639 -- a confirmed visual binary that contains
an eclipsing component, which is not detected in the spectra. 
We present strong clues that the latter system may, however, host additional,
sub-stellar body.

\subsubsection{KIC~08718273 -- a blend with a foreground giant}

This system has been observed nine times with HIDES by us, and twice by the APOGEE
survey. We noted only one set of lines, whose measured velocity occurred to be relatively 
stable, with $rms=68$~\ms, which was slightly more than for the RV standards. 
Phase-folding with period from the KEBC of 6.959~d, or half of this value,
and fitting a sine function, did not lead to any improvement. This confirmed
the conclusions of \citet{gau13} that the oscillations they found are
coming from a red giant not related to the eclipsing pair. We thus observe 
a blend. We perform a fit to the RVs in order to detected a signal that could 
be induced by the oscillations, but the details are described in Section \ref{sec_rvosc}.
Due to a relatively large number of degrees of freedom, we used this fit to look for 
the difference between HIDES (plus ATLAS~9/12 templates) and APOGEE RV systems, 
which we found to be 112$\pm$22~\ms. We used it to shift APOGEE measurements of 
other targets, and included its uncertainty to the individual measurement errors.
This case also confirms the correctness of the instrumental correction we found for the 
December 2014 run. Furthermore, we have found no trend indicating orbital motion around 
the centre of mass common with the eclipsing pair, confirming previous 
conclusions.

\subsubsection{KIC~05598639 -- a possible multiple}

\begin{figure}
\centering
\includegraphics[width=0.9\columnwidth]{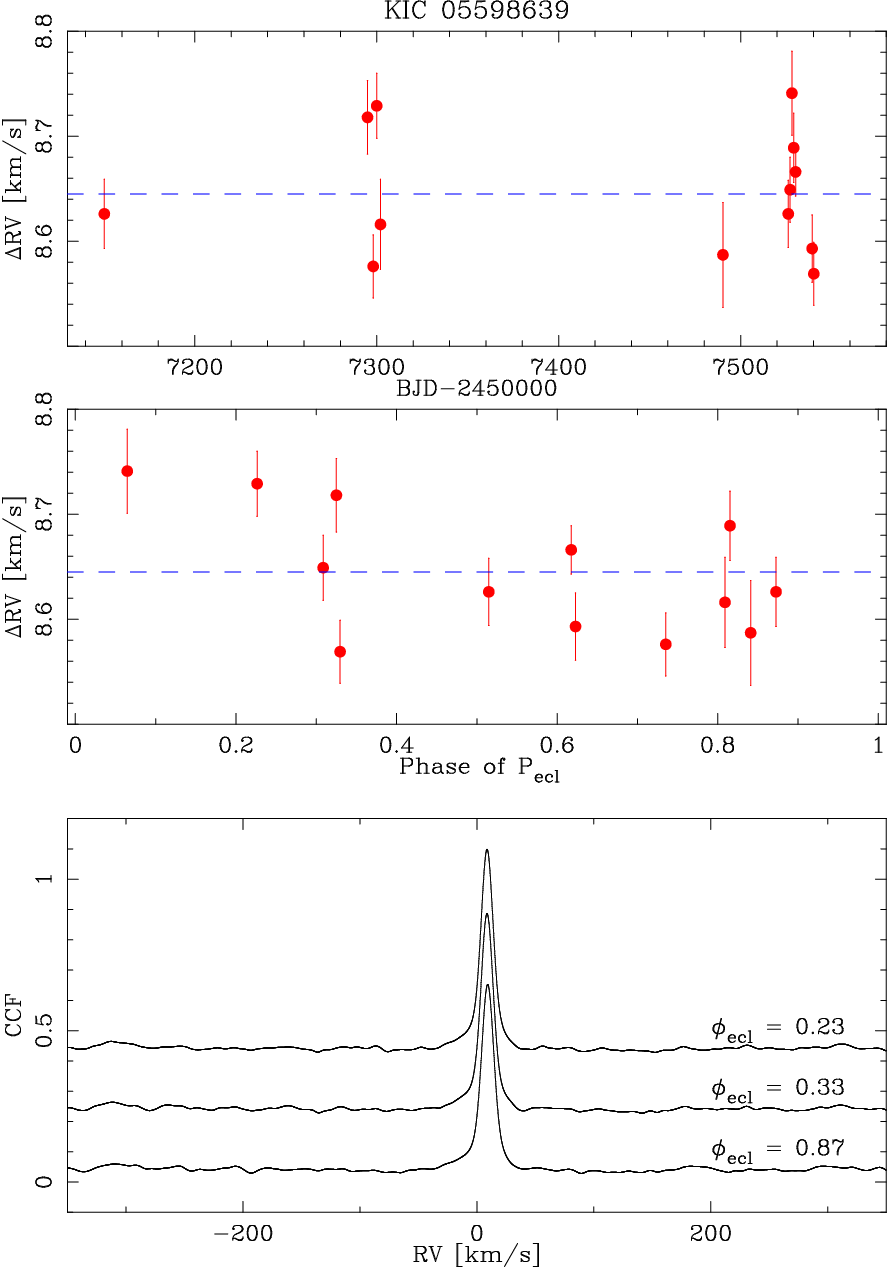}
\caption{{\it Top:} Our HIDES radial velocities of KIC~05598639 as a function of
time. The blue line marks the average measured velocity. The $rms$ is 59 \ms. 
{\it Middle:} Same, but phase-folded with the ephemeris of the eclipsing pair. 
No change related to the KEBC ephemeris is seen.
{\it Bottom:} Three cross-correlation functions (CCFs), for spectra taken
during different phases of the eclipsing binary $\phi_{ecl}$, shifted for clarity.
Only one narrow peak is seen, and no others, that could
be coming from the eclipsing binary, can be identified.
}\label{fig_rv_0559}
\end{figure}

\begin{figure}
\centering
\includegraphics[width=0.9\columnwidth]{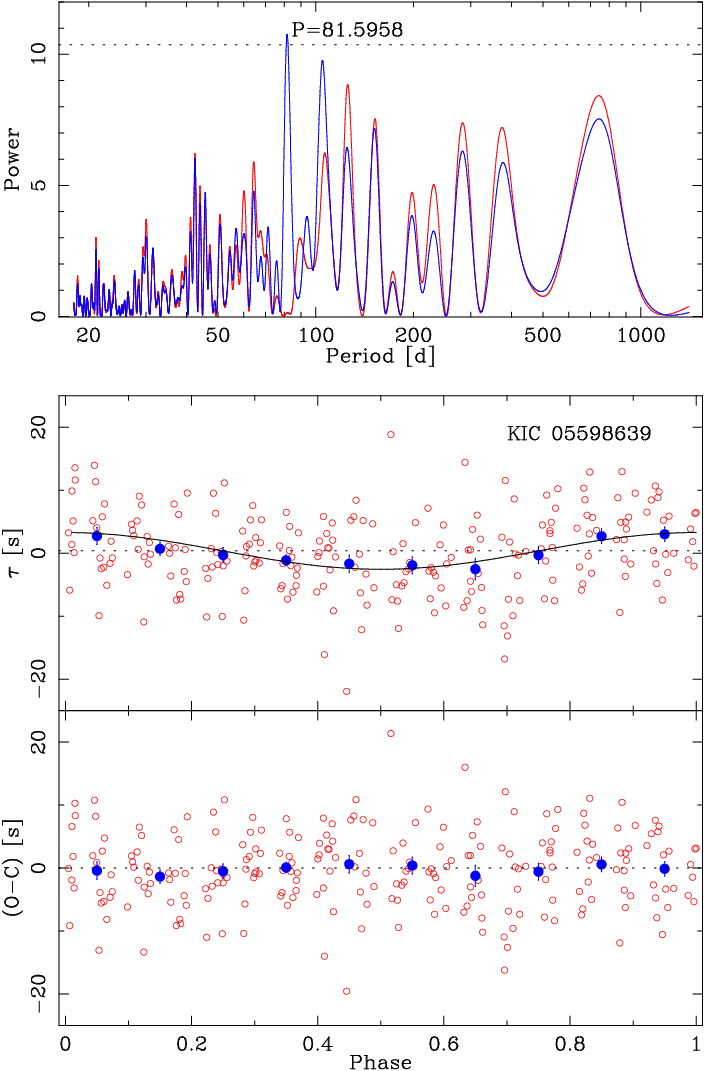}
\caption{{\it Top:} Periodograms of our ETV measurements of KIC~05598639. 
The blue line is for the original data, for which the highest peak at 81.6 days
is marked. The red line is for the data after this period is removed.
The dotted line marks the FAP of 0.5\%.
{\it Middle:} Our ETV measurements phase-folded with the period of 81.6 days.
Red points are original data, and blue points are data binned over 0.1 in
orbital phase. 
The black solid line is the best cosine fit to the ETVs, and the dotted line is set at 
zero, for comparison. {\it Bottom:} Residuals of the best cosine fit to the ETVs.  
}\label{fig_etv_0559}
\end{figure}

To date we have obtained thirteen HIDES observations of this system, and were not
able to find any other spectroscopic data. The light curve shows two wide, equal-depth
eclipses and relatively strong ellipsoidal variations. Assuming that the eclipsing pair
is composed of two solar-analogues, and taking $P_{ecl}=1.29755$~d from KEBC, we 
would expect to see two sets of equal lines, rotationally widened to $\sim$40~k\ms\,
($\propto R$), showing velocity variation with semi-amplitude of $>$120~k\ms\,
($\propto M^{1/3}$). Meanwhile, we only see one set of narrow lines, that 
is relatively stable (Fig.~\ref{fig_rv_0559}). 
The \jkt solution suggests that the eclipsing pair contributes only 38 per cent to 
the total flux, so the lines seen in the spectra probably come from the brighter 
component of the visual pair, and the fainter is the eclipsing binary 
observed by {\it Kepler}. The resulting magnitude difference -- 0.53~mag -- 
is not in agreement with the WDS value. 
The measured RVs are stable down to $\sim$60~\ms, which is the level 
of the $rm$ of RV standards. We therefore conclude that no significant motion
can be detected in our data.

The ETVs calculated by \citet{con14} seem to be affected by strong systematics, so 
we have calculated our own, as in \ref{sec_etv}. We found no obvious 
periodicity at first, but we run a Lomb-Scargle periodogram on our 
measurements\footnote{Periodograms for this work were 
created with the on-line NASA Exoplanet Archive Periodogram Service:\\ 
{\tt http://exoplanetarchive.ipac.caltech.edu/cgi-bin/Pgram/nph-pgram}.}. We searched
for periods between 18 and 1433 days (the time span $W$ of the data), using fixed 
frequency step of $1/10W = 6.98\times10^{-5}$~d$^{-1}$. The results are shown 
in Figure~\ref{fig_etv_0559}.

The highest peak of the periodogram was found at 81.6 days, but it was not much
higher than the second one at 104.86~d. We fitted a cosine function with the 81.6~d
period, and found the amplitude $A$ of $2.71\pm0.56$~s (formal uncertainty of the fit
only). This fit is also shown in Figure~\ref{fig_etv_0559}. To show the change
of the ETVs with more clarity, we also plot data binned over 0.1 in phase. 
At the given period, the amplitude corresponds to minimum mass function
of the outer body of $f(m)=3.21(66)\times10^{-6}$~M$_\odot$,
or $3.36(69)\times10^{-3}$~M$_{JUP}$. Assuming that the total mass of the inner
binary $M_{tot}$ is 2~M$_\odot$, and the outer body is significantly
less massive than the eclipsing binary, its lower mass limit is then 
$m \sin(i) = 25(1)$~M$_{JUP}$ ($\propto M_{tot}^{2/3}$). We would like to note
that $M_{tot}$ may actually be around 1~M$_\odot$, or less. \citet{hub14}
estimated the mass of the dominant component (the one we see in spectra) 
to be $0.796^{+0.112}_{-0.073}$~M$_\odot$. The components of 
the inner binary should then be even less massive, possibly below 0.5~M$_\odot$
each. The $M_{tot}$ of 1~M$_\odot$ gives $m \sin(i) \sim 15$~M$_{JUP}$, pushing
the outer body's lower mass close to the planetary value. However,
due to lack of direct information about the eclipsing pair, and large uncertainties of 
values from \citeauthor{hub14} and our eclipse timing solutions, we find this
possibility rather unlikely.

The eclipse timing amplitude we found is 2.2 times smaller than the $rms$ of 
the fit (6.018~s), thus we can not conclude that this periodicity is real, but is 
detected at 4.8$\sigma$ level ($\Delta A/A$), and we estimate the false-alarm 
probability to about 0.36 per cent. We also note that on the periodogram
of the residuals (shown in Fig.~\ref{fig_etv_0559} as the red line), the 
peak at 104.86~d is not the most prominent one. If the 81.6~d period of 
the eclipsing pair is real, the system may be a hierarchical quadruple, bearing a
circumbinary brown dwarf, or even a massive planet. To confirm that, one needs more 
precise eclipse timing measurements and more spectroscopic observations. 
The true character of KIC~05598639 remains unknown, and more observations 
are planned.

\subsection{RV motion not coinciding with $P_{ecl}$}

In this section we present the analysis of two cases --
 KIC~04758368 and 10991989 -- where the observed RV variations
do not coincide with the period of eclipses seen in the \kep data.
They are hierarchical triple systems with the third (outer) star dominating 
the flux and seen in the spectra. They were treated as double-lined spectroscopic 
binaries, and parameters of their outer orbits were derived. 
The RVs of the centre of mass of the inner pair were obtained by translating the eclipse 
timing variations, as described in Section \ref{sec_oc2rv}, but they contain limited 
information on the systemic velocity of the common centre of mass (CCM) $\gamma$. 
In order to find it, the inner pair was treated in the \vfit code as the primary
and $\gamma_1$ was fitted for. The outer star (dominant in flux) was the secondary, 
and $\Delta\gamma$ has been found. The true systemic velocity of the CCM 
is thus $\Delta\gamma-\gamma_1$.

Although it may be the dominant source of flux in the triple, the outer star is 
always designated as the component B, and the inner eclipsing binary as A (Aa+Ab). 
The results are summarised in Table \ref{tab_par_3rd}. 
As previously, uncertainties include systematics calculated with a bootstrap routine. 
Among other parameters, we show times of the pericentre passage $T_{per}$, number of 
data points for each component $N_{A,B}$ (note there are many more from ETVs than 
from direct RVs, but are much more spread), and the pericentre longitude given for 
the inner binary. We also give the systemic velocity of the whole triple as $\gamma$.

Moreover, for these two targets the \jkt fits were performed to estimate the 
contribution of each component to the total flux, inclination of the 
inner orbit, and fractional radii of the inner components -- $R_{Aa}/a_A$ and
$R_{Ab}/a_A$. Using the literature estimates of mass of the outer star $M_B$, we 
calculate the inclination of the outer orbit $i_{AB}$, total mass of the inner
pair $M_A$, its major semi-axis $a_A$, and absolute radii $R_{Aa}$ and $R_{Ab}$.

\begin{table}
\centering
\caption{Orbital and stellar parameters of two hierarchical triples showing
lines of the outer star in their spectra.}\label{tab_par_3rd}
\begin{tabular}{lcc}
\hline \hline
KIC			& 04758368 & 10991989 \\
\hline
{\it Outer orbit} &&\\
$P_{AB}$~(d)		& 2553(80) & 547.81(56) \\
$T_0$ (BJD-2450000)	& 5581(9)  & 5490(2) \\
$K_A$ (k\ms)		& 8.81(41) & 10.28(13) \\
$K_B$ (k\ms)		&12.9(1.4) &  7.91(11) \\
$e_{AB}$		& 0.672(37)& 0.251(7) \\
$\omega_A$ ($^\circ$)	&141.9(2.1)& 18.6(1.2) \\
$\gamma$ (k\ms)	&-24.0(1.2)& -18.78(16) \\
$M_A \sin^3(i)$ (M$_\odot$)	& 0.65(18)& 0.135(4)\\
$M_B \sin^3(i)$ (M$_\odot$)	& 0.45(9) & 0.175(5)\\
$a_{AB}\sin(i)$ (AU)	&3.77(32)& 0.886(8)\\
$N_A$		& 106	& 101	\\
$N_B$		& 19	& 14	\\
$rms_A$ (k\ms)	& 6.9	& 1.22	\\
$rms_B$ (\ms)	& 57    & 25$^a$ \\
\hline
\jkt {\it solution} &&\\
$r_{Aa} + r_{Ab}$	& 0.4252(64) & 0.373(10) \\
$r_{Ab}/r_{Aa}$  	& 1.22(21)   & 0.657(58) \\
$L_B/L_{tot}$		& 0.725(19)  & 0.9764(54)\\
$i_A$ ($^\circ$)	& 73.94(79)  & 85.6(1.5)\\
\hline
{\it Absolute values} &&\\
$M_B$ (M$_\odot$)	&  1.43 / 0.85$_{-0.02}^{+0.43\,b}$	& 2.5(4)$^c$\\
$i_{AB}$ ($^\circ$)	&    43 / 53.7$_{-12.3}^{+5.7}$		& 24.3(1.4)\\
$M_A$ (M$_\odot$)	&   2.1 / 1.24$_{-0.15}^{+0.65}$	& 1.92(31)\\
$a_{AB}$ (AU)		&   5.3 / 4.68$_{-0.67}^{+1.50}$	& 2.15(12)\\
$a_A$ (R$_\odot$)	&  13.0 / 10.92$_{-1.56}^{+3.50}$	& 5.14(28)\\
$R_{Aa}$ (R$_\odot$)	&2.48(30) / 2.09$_{-0.39}^{+0.72}$ 	& 1.16(7) \\
$R_{Ab}$ (R$_\odot$)	&3.04(23) / 2.56$_{-0.42}^{+0.84}$	& 0.81(6) \\
\hline
\end{tabular}
\\$^a$Before fitting for the oscillations.
\\$^b$From the KIC (without errors) and \citet{hub14}.
\\$^c$From \citet{gau13}.
\end{table}

\subsubsection{KIC~04758368}
\begin{figure}
\centering
\includegraphics[width=0.9\columnwidth]{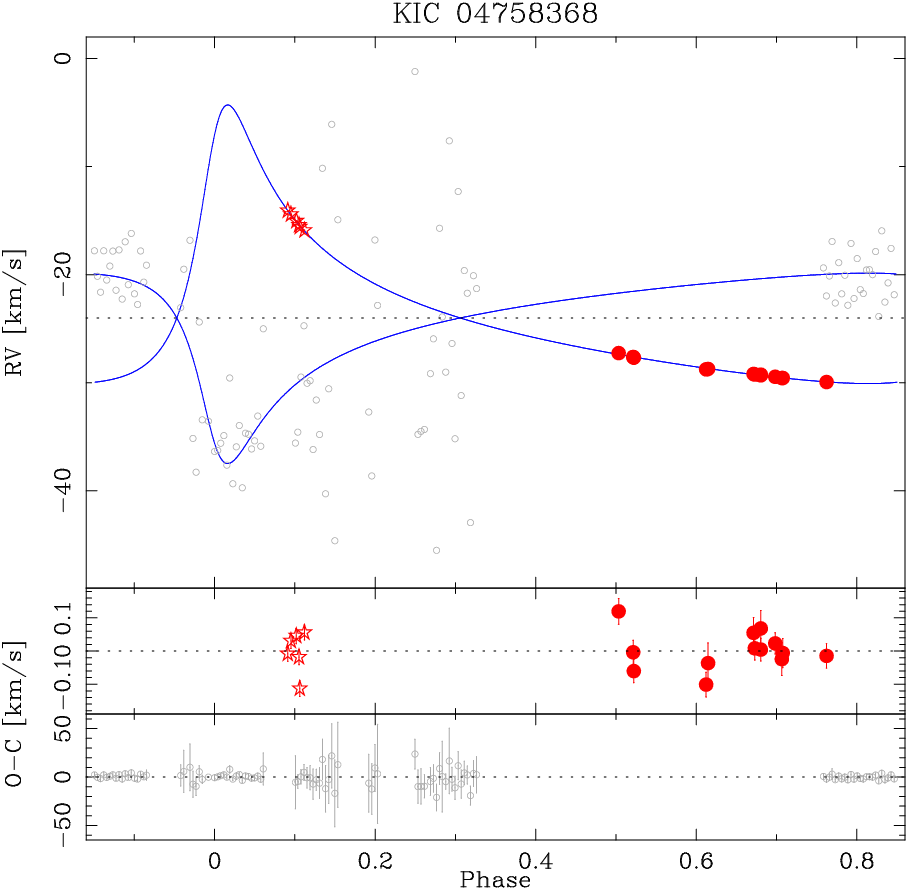}
\caption{Orbital solution of the outer orbit of KIC~04758368. Grey
points show the RVs of the centre of mass of the inner eclipsing
binary, reconstructed from the ETV measurements,
and shifted to the systemic velocity of the whole triple (dotted line).
Red points are the direct RV measurements of the outer star, taken
from the APOGEE data base (opens stars), and from our HIDES 
spectra (filled circles). Blue lines show the best-fitting orbital 
solution. Phase 0 is set to the moment of the pericentre passage.
Lower panels show the residuals of the fit, separately for
the direct and post-ETV measurements.}\label{fig_orb_0475}
\end{figure}
\begin{figure}
\centering
\includegraphics[width=0.9\columnwidth]{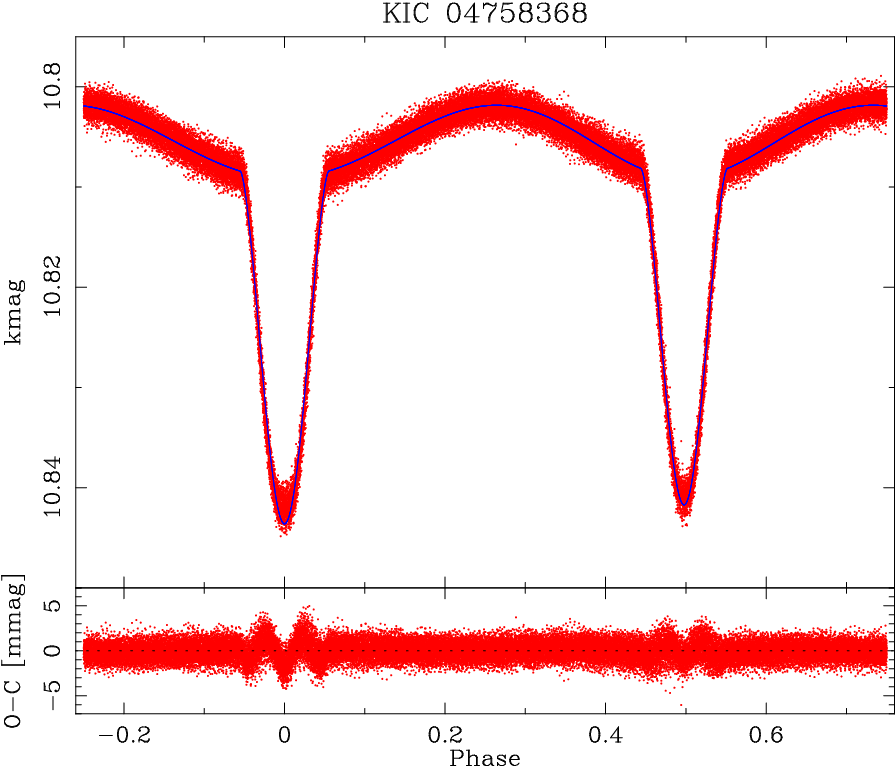}
\caption{\jkt light curve solution for KIC~04758368.}
\label{fig_lc_0475}
\end{figure}

The very long outer period of this system (the longest in the sample) makes
it very difficult to analyse with the currently available data. The 
ETV measurements do not cover it completely, but their distinct maximum 
around JD$\sim$2455400, which means a sudden change in radial velocity, 
suggests that the inner binary passed the outer orbit's pericentre that 
time (Fig.~\ref{fig_oc2rv}). \citet{con14} do not give the complete orbital 
solution, and the one by \citet{bor16} is marked as uncertain.

Our thirteen HIDES RV measurements cover even smaller 
fraction of the period, so the APOGEE data were very important. 
Initially, the post-ETV data were showing a best-fitting period of 
about 1600 days, which for the May 2015 (JD$\sim$2457150) observations 
predicted orbital phases roughly around the pericentre, i.e. RVs significantly
higher than obtained from the previous HIDES runs. However, the measured velocity
was still decreasing a year later -- in May 2016 (JD$\sim$2457527), 
meaning that the star still did not pass the periastron, and that the period is 
actually longer than  $\sim$2200 days. We still don't have enough data 
to securely give its value, and we treat our solution as preliminary.

The best-fitting orbital model is shown in Figure~\ref{fig_orb_0475},
and corresponding parameters in Table \ref{tab_par_3rd}.
The main source of uncertainty is the orbital period, which may still 
be longer than we give, and its uncertainty underestimated. Another one
are the errors and scatter of post-ETV velocities, mainly between phases 
0.1 and 0.4, where the ETVs show a small curvature, comparable to
or smaller than their scatter (Fig.~\ref{fig_oc2rv}).

We found that the third star is less massive than the inner binary, 
and constitutes about 72 per cent of the total flux. 
\citet{gau13} identify it as a non-oscillating red giant, and two 
estimates of mass/radius/gravity for this star can be found in catalogues.
The MAST archive gives $\log(g_B)=2.624\pm0.400$ (from the KIC) 
and $R_B=9.651$~R$_\odot$ \citep[without errors, from][]{chr12}, 
which translates into 1.43~$M_\odot$. \citet{hub14} give the revised 
$\log(g_B)=2.610^{+0.261}_{-0.254}$, and estimates $R_B=7.55^{+3.49}_{-2.14}$~R$_\odot$ 
and $M_B=0.85^{+0.43}_{-0.02}$~M$_\odot$ ($\pm68$ per cent confidence level). 

In Table \ref{tab_par_3rd} we also show the
absolute mass of the inner pair and parameters of the outer orbit, assuming the two
values of $M_B$. We note that the \kep light curve (Figure~\ref{fig_lc_0475}) 
shows nearly equal eclipses, 
suggesting nearly identical parameters of the components, and substantial
ellipsoidal variations, suggesting their large size, relatively to their 
physical separation. The period of the inner orbit is 3.75~d, so we find it 
plausible that the components Aa and Ab are actually sub-giants, with masses
comparable to or higher than the Sun, only significantly larger
This is in agreement with the estimates of their absolute radii.
All this favours the physical parameters of the star B found in the MAST archive, 
or around the upper limit of Huber's values. The whole system would thus be
relatively old, which agrees with the metal abundance estimate [$M/H$]=-0.46~dex
\citep[from both KIC and][]{hub14}. However, as it will be discussed later
in this work, the MAST values of $\log(g)$ may be overestimated.

We would also like to note that in this case the APOGEE archive gives very different 
values of $log(g)$ and [$M/H$] of 2.24(11) and 0.01(3)~dex, respectively. 
It is possibly because the analysis was done on an averaged spectrum, made 
of six visit spectra, each having different heliocentric velocity. 
Averaging such spectra, without a prior velocity correction, might have 
resulted in a spectrum with lines wider than they really are.

\subsubsection{KIC~10991989}

\begin{figure}
\centering
\includegraphics[width=0.9\columnwidth]{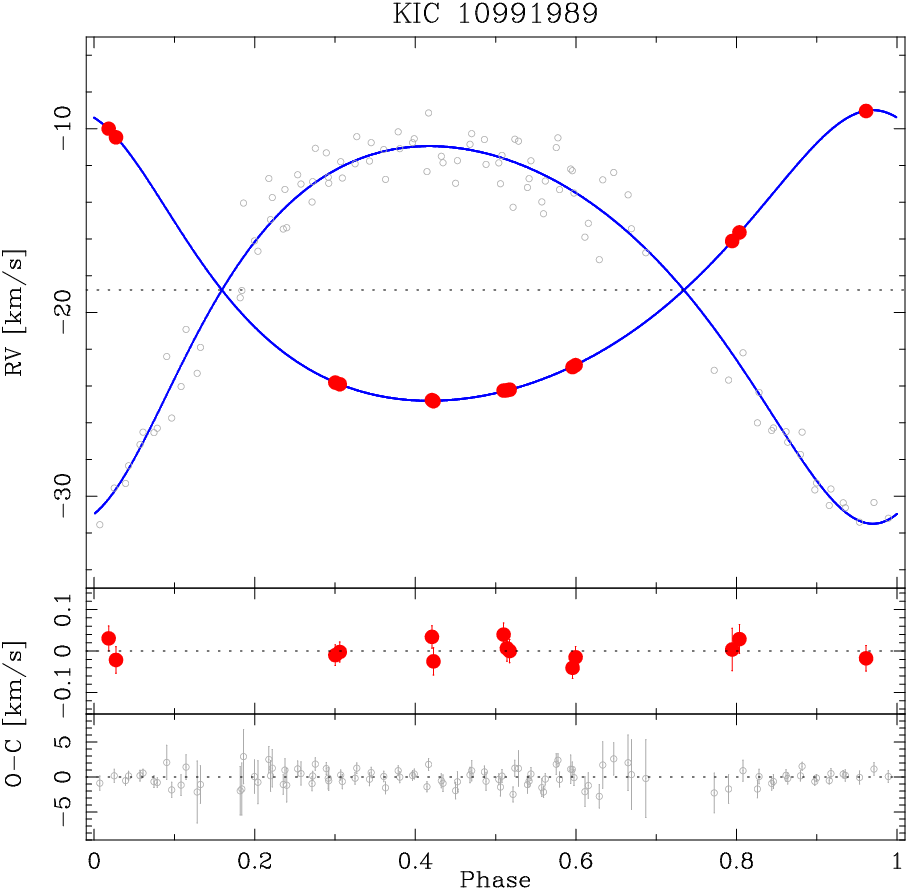}
\caption{Same as Fig.~\ref{fig_orb_0475}, but for KIC~10991989.
All directly measured RVs (red points) come from our HIDES spectra.
}\label{fig_orb_1099}
\end{figure}
\begin{figure}
\centering
\includegraphics[width=0.9\columnwidth]{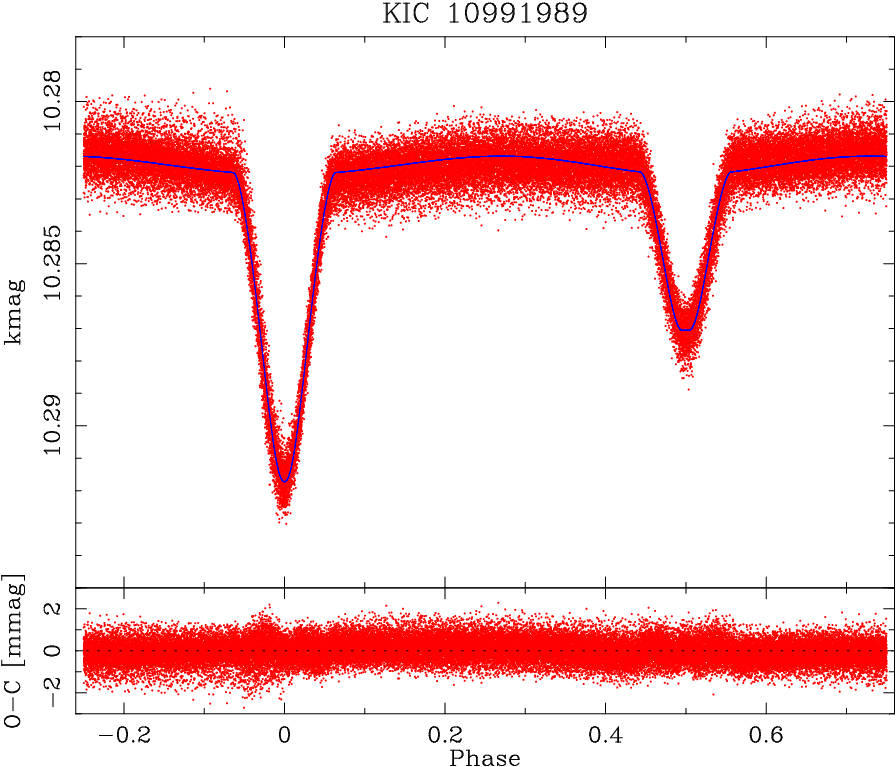}
\caption{Same as Fig.~\ref{fig_lc_0475}, but for KIC~10991989.
}\label{fig_lc_1099}
\end{figure}

We have observed this system fourteen times with HIDES, covering the period 
of the outer orbit almost completely, and the ETVs cover almost 2.5 cycles 
(Fig.~\ref{fig_oc2rv}). Our solution is depicted in Figure~\ref{fig_orb_1099}
with parameters summarized in Table \ref{tab_par_3rd}.
Our results for the outer orbit agree with those of \citet{rap13} in terms of 
orbital parameters, i.e. $e_{AB},\omega_{A}, M_{A,B}\sin^3(i)$, and the mass 
ratio $M_B/(M_A+M_B)$, however their uncertainties are quite large.
A slight disagreement is seen in ephemeris -- $P_{AB}$=547.81(56) vs. 554.2~d, 
and $T_{per}=2455490(2)$ vs. $2455525.7^{+21}_{-18}$ -- although 
\citeauthor{rap13} do not give the uncertainty of their outer period.
The outer period in the solution of \citet{bor16} is in a much better agreement 
with ours -- 548(1)~d -- but their $e$ and $\omega$ are not, being
0.35(2) and 29(4)$^\circ$, respectively.
We find another solution, by \citet{con14}, less reliable, because it assumes a
circular orbit, while our direct RVs show a substantial eccentricity. 

We also used the component B (pulsating red giant) mass estimate from
\citet{gau13}, and calculated the total mass of the inner binary, 
and outer orbit's inclination. By combining this with the results from \jkt run
(Figure~\ref{fig_lc_1099}),
we can estimate the inner orbit's major semi-axis, and absolute radii of Aa and Ab.
The results are also shown in Table \ref{tab_par_3rd}.
Additionally, we estimate (from the \jkt analysis) that the outer star
contributes 97.6 per cent to the total flux. The inner binary is 
composed of two stars of significantly different radii 
(note that the secondary eclipse is total), and different spectral 
types (as can be deduced from unequal eclipses, for example). 
Considering the resulted radii $R_{Aa}$ and $R_{Ab}$,
it's possible that the inner pair is composed of $\sim$1.15+0.75~M$_\odot$ 
main sequence stars. At least one component shows a substantial
level of activity -- single-quarter \kep light curves reveal out-of-eclipse
modulation coming from cold spots that slowly evolve in time. 
The fact that the brightness of the system in both minima varies 
with the same scale, suggests similar influence of spots on the flux 
of both stars. This means that the secondary, which is likely much smaller 
and cooler than the primary, must have relatively larger area covered in 
spots.

\subsection{RV motion coinciding with $P_{ecl}$}

In this section we present results for five single-lined spectroscopic binaries
(SB1), i.e. systems for which the period of the RVs is the same as of the eclipses. With the 
exception of KIC~10614012, these systems show both eclipses in the \kep data. For those
we performed a \jkt analysis to find the inclination, fractional radii, brightness ratio,
as well as period, eccentricity and longitude of pericentre. Last three parameters were
used in {\sc v2fit}, with which we found the systemic velocity, velocity amplitude, moment of
the periastron passage or quadrature (for $e=0$), and the major semi-axis of the primary's
orbit around the centre of mass. These results are summarised in Table~\ref{tab_par_1001}.

Further comparison with isochrones was also made to estimate the absolute values of
stellar parameters, age, and distance. These results are shown in Table~\ref{tab_1001_phys}.

\begin{table*}
\centering
\caption{\vfit and \jkt solutions for KIC~03120320, 08430105, 10001167, and 10015516.
}\label{tab_par_1001}
\begin{tabular}{lcccc}
\hline \hline
Parameter 			& 03210320 	& 08430105 	& 10001167 	& 10015516 \\
\hline
$P$ (d)				& 10.2656136(4) & 63.32717(5)	& 120.39110(26)	& 67.692165(15) \\
$T_{0}$ (BJD-2450000)$^a$ 	& 4956.29(26)	& 4985.91(42)	& 5064.0(1.1) 	& 5022.291(8) \\
$T_{ecl}$ (BJD-2450000)$^b$	& 4959.22316(3)	& 4976.6345(7)	& 4957.6827(19)	& 5005.38019(18) \\
$e$				& 0.034(5) 	& 0.256(72) 	& 0.160(2) 	& 0.0(fix) \\
$\omega$ ($^\circ$) 		& 343(9) 	& 350.5(1.6) 	& 213(1) 	& --- \\
$K_1$ (k\ms) 			& 39.23(9) 	& 27.66(13) 	& 24.9(7) 	& 32.162(54) \\
$\gamma$ (k\ms) 		& -41.10(26) 	& 11.5(1.0) 	& -103.2(1.1) 	& -16.471(37) \\
$a_1 \sin(i)$ (AU) 		& 0.03700(24) 	& 0.1556(8) 	& 0.272(8) 	& 0.20012(34) \\
$a_1 \sin(i)$ (R$_\odot$) 	& 7.956(2)	& 33.47(16)	& 58.5(1.7) 	& 43.03(7) \\
$f(M_2)$ (M$_\odot$) 		& 0.06411(44) 	& 0.1254(74)	& 0.185(16) 	& 0.2333(12) \\
$r_1$				& 0.05898(44) 	& 0.0893(21) 	& 0.112(18)  	& 0.0982(1) \\
$r_2$				& 0.02287(65) 	& 0.00905(37) 	& 0.0085(20) 	& 0.01536(20) \\
$i$ ($^\circ$) 		& 87.845(73) 	& 88.53(53) 	& 86.6(1.9) 	& 86.85(10) \\
$L_2/L_1$ 			& 0.0127(7) 	& 0.01729(46) 	& 0.0189(9) 	& 0.1046(20) \\
$rms_{RV}$ (\ms) 		& 153		& 71   		& 137 		& 34 \\
$rms_{LC}$ (mmag) 		& 0.92 		& 0.66 		& 1.3 		& 0.69 \\
\hline
\end{tabular}
\\$^a$ Time of periastron passage ($e>0$) or quadrature ($e=0$).
\\$^b$ Mid-time of primary (deeper) eclipse.
\end{table*}

\begin{table*}
\centering
\caption{Results of the isochrone mass/age search for 
KIC~03120320, 08430105, 10001167, and 10015516.}\label{tab_1001_phys}
\begin{tabular}{lcccccccc}
\hline \hline
Parameter & \multicolumn{2}{c}{03120320} & \multicolumn{2}{c}{0843105} & \multicolumn{2}{c}{10001167} & \multicolumn{2}{c}{10015516}\\
\hline
	&\jkt &{\it Isochrones} &\jkt &{\it Isochrones}
	&\jkt &{\it Isochrones} &\jkt &{\it Isochrones}\\
$r_1 + r_2$	& 0.08186(66)& 0.08178 & 0.0985(35) & 0.01021 & 0.121(19) & 0.122  & 0.1136(10) & 0.1128 \\
$r_2 / r_1$	& 0.388(9)   & 0.4122  & 0.1006(25) & 0.1016  & 0.0764(65)& 0.0757 & 0.1563(22) & 0.1564 \\
$L_2 / L_1$	& 0.0127(7)  & 0.0123  & 0.01729(46)& 0.01716 & 0.0189(9) & 0.0190 & 0.1046(20) & 0.1038 \\
\hline
\multicolumn{2}{l}{\it Individual, absolute}\\
	& {\it Primary} & {\it Secondary} & {\it Primary} & {\it Secondary} 
	& {\it Primary} & {\it Secondary} & {\it Primary} & {\it Secondary} \\
$M$ (M$_\odot$)& 1.019 & 0.550 & 1.042 & 0.790	& 1.135 & 0.993 & 2.024 & 1.480 \\
$R$ (R$_\odot$)& 1.338 & 0.552 & 7.582 & 0.770	&14.984 & 1.134 &10.363 & 1.620 \\
$\log(g)$	& 3.926 & 4.696 & 2.576 & 4.562 & 2.084 & 4.326 & 2.577 & 4.189 \\
$T_{eff}$ (K)	& 5780  & 3945  & 4890  & 5790  & 4750  & 6430  & 4960  & 7610  \\
$Kmag$ (mag)	& 3.430 & 8.204 & 0.905 & 5.319 &-0.269 & 4.034 & 0.081 & 2.527 \\
$K$ (k\ms)$^a$	& 39.23 & 73.1  & 27.7  & 33.47 & 24.9  & 28.4  & 32.16 & 42.39 \\
\hline
\multicolumn{2}{l}{\it Whole system}\\
$\left[M/H\right]^b$	& \multicolumn{2}{c}{-0.24}& \multicolumn{2}{c}{-0.60}& \multicolumn{2}{c}{-0.50}& \multicolumn{2}{c}{-0.22} \\
$t$ (Gyr)		& \multicolumn{2}{c}{7.94} & \multicolumn{2}{c}{6.31} & \multicolumn{2}{c}{4.47} & \multicolumn{2}{c}{1.26} \\
$d$ (kpc)$^c$		& \multicolumn{2}{c}{0.31} & \multicolumn{2}{c}{0.81} & \multicolumn{2}{c}{1.17} & \multicolumn{2}{c}{1.40} \\
\hline
\end{tabular}
\\$^a$Directly from Tab.~\ref{tab_par_1001} for the primary.
$^b$From MAST.
$^c$Assuming no interstellar extinction.
\end{table*}

\subsubsection{KIC~03120320}
\begin{figure*}
\centering
\includegraphics[width=0.9\columnwidth]{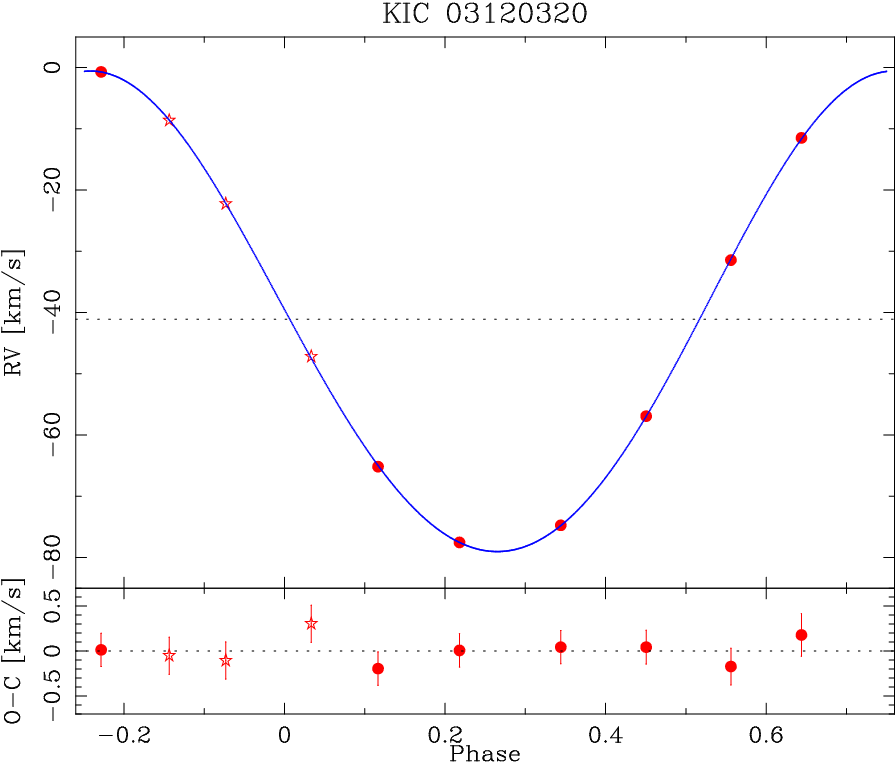}
\includegraphics[width=0.9\columnwidth]{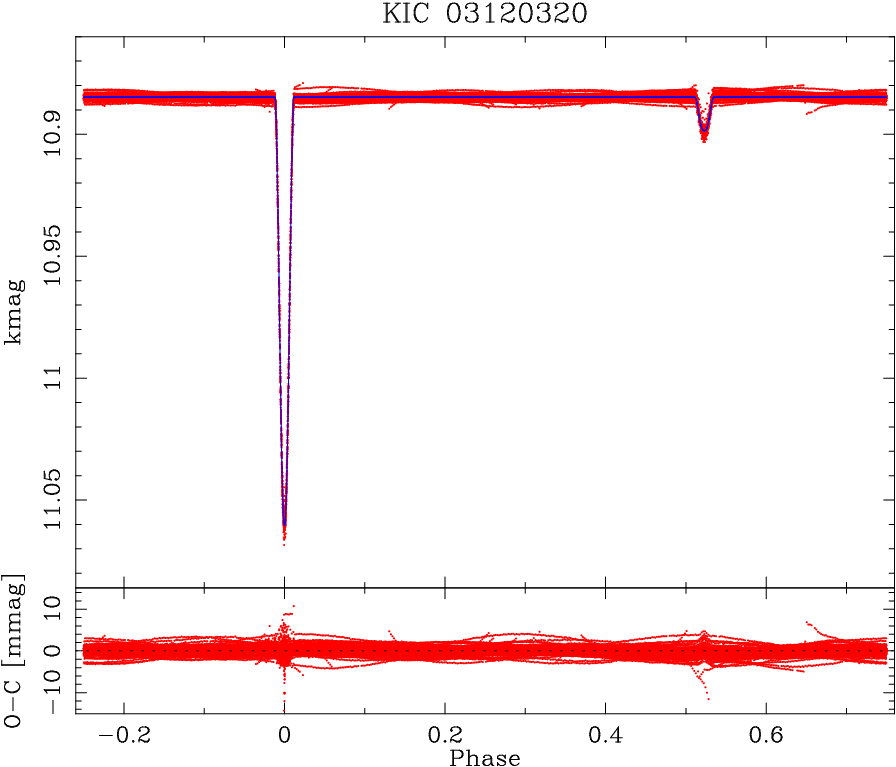}
\caption{{\it Left}: Orbital fit (blue line) to our HIDES (filled circles)
and APOGEE (open stars) RVs of KIC~03120320, phase-folded with the ephemeris from 
Table~\ref{tab_par_1001} ($P,T_{ecl}$). The decrease of the velocity during phase 0 
(deeper eclipse) means that the RVs are measured for the hotter component. {\it Right}
\jkt fit the the complete Q0-Q17 curve.
}\label{fig_orb_0312}
\end{figure*}

This target has been observed seven times with HIDES by us, 
and three more by the APOGEE survey. Among the four SB1s analysed 
with \jkt this is the only one where the primary eclipse is not 
total, and the component visible in the spectra is the hotter one. 
This can be deduced from the fact that its RV decreases during the primary
eclipse. The secondary one is not at the phase 0.5, meaning that the orbit is 
slightly elliptical. Also, the fit to the RVs is slightly better for a non-circular
orbit, than for $e=0$. Results of our fits are summarised in Table~\ref{tab_par_1001}, 
and on Figure~\ref{fig_orb_0312}. We note an out-of-eclipse modulation, most likely
coming from the presence of spots on the primary. Its influence on the RV measurements
may explain the relatively large $rms$ of the orbital fit.

On the [$M/H$]=-0.24~dex PARSEC isochrone, we found a good 
match to \jkt values for a $1.019+0.550$~M$_\odot$ pair at the age of almost 8~Gyr 
($\log(t)=9.90$). The primary is somewhat evolved, with the radius of $1.338$~R$_\odot$,
and entered the sub-giant phase, while the secondary is on the main sequence 
($R_2=0.552$~R$_\odot$). Its lines
should be detectable on very high $S/N$ ratio spectra, especially in infra-red.
The secondary should be highly active, and its emission lines may be seen.
Unfortunately, HIDES does not reach to the calcium H and K lines, where such emission
would be easiest to note. Assuming no interstellar extinction, we estimate the 
distance to this pair to 311~pc, thus it should be measurable by the {\it Gaia} satellite.

We would like to note that even better match was found for a $0.932+0.517$~M$_\odot$
pair at the age of 11.2~Gyr ($\log(t)=10.05$). However, we found this solution
less reliable, as the predicted $g'-r'$ colour is 0.554~mag, which is significantly
redder than the observed 0.453. The solution we adopted predicts $g'-r'=0.440$~mag, 
and the difference can be attributed to the reddening, or uncertainties of \jkt 
results or the metallicity. It is worse in terms of $\chi^2$ mainly due to the 
disagreement in ratio of the radii $k$. This is because, when both eclipses are partial, 
this parameter is not constrained very well by the light curve only. The usual 
way of solving this issue is by using spectroscopy-based flux ratios. This is, however, 
impossible when lines of the secondary are not visible in the spectra. We suspect 
that our uncertainties in $k$ (and $L_2/L_1$) for this system may be underestimated, 
but still the agreement with the isochrone values is better than 3$\sigma$.

\subsubsection{KIC~08430105}
\begin{figure*}
\centering
\includegraphics[width=0.9\columnwidth]{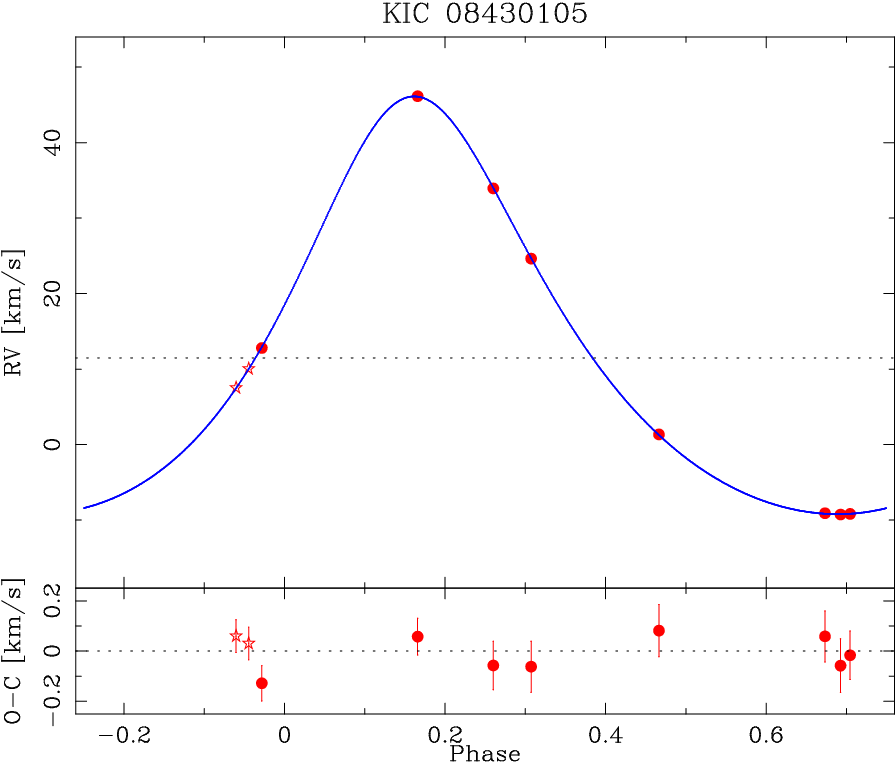}
\includegraphics[width=0.9\columnwidth]{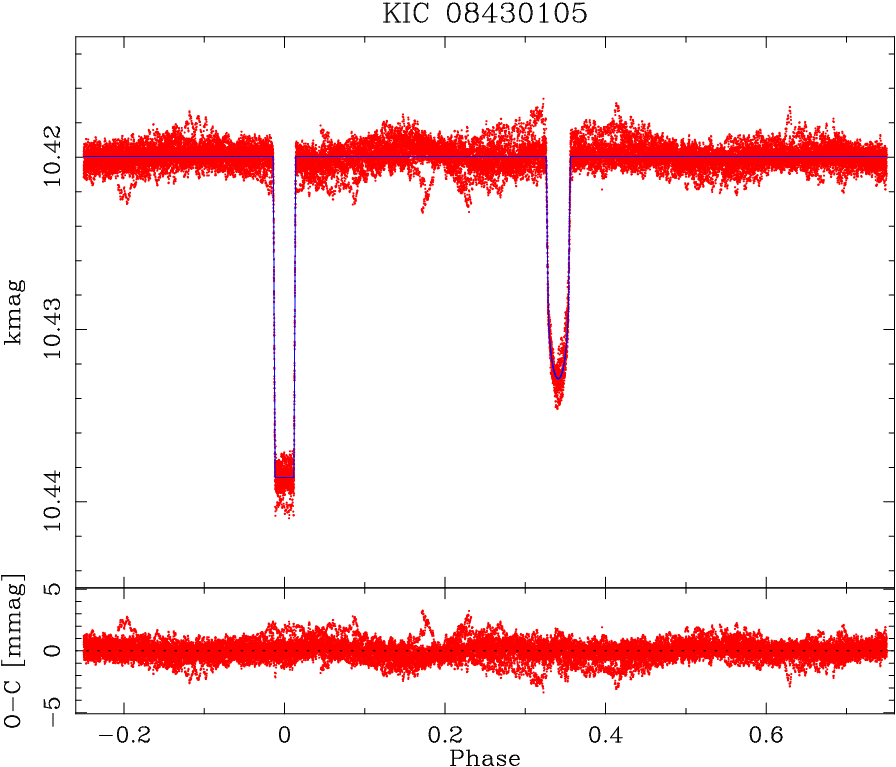}
\caption{Same as Fig.~\ref{fig_orb_0312} but 
for KIC~08430105. The increase of the velocity (left) during phase 0 
(deeper eclipse) means that the RVs are measured for the cooler 
component.
}\label{fig_orb_0843}
\end{figure*}

It is the first of three SB1 systems in our sample that show a total primary 
eclipse, which helps a lot in the LC analysis. It is also the most metal-poor
among them, with [$M/H$]=-0.596~dex (from MAST). We have acquired eight 
spectra with HIDES, and found two more observations in the APOGEE archive.
The data points are spread more-less evenly throughout the orbital period,
including phases around maximum and minimum of the RV curve
therefore the RV-based parameters -- $T_{per}, K_1$, and  $\gamma$ -- are 
constrained quite well. The final $rms=71$~\ms\, is comparable to those 
of the RV standards. Also, $e$ and $\omega$ can be estimated 
precisely from the light curve. Results of our fits are summarised
in Table~\ref{tab_par_1001}, and on Figure~\ref{fig_orb_0843}.

From the fact that the RV curve rises
during the total primary eclipse, we can deduce that the primary is cooler
and larger than the secondary. Indeed, comparison with the PARSEC isochrone
for [$M/H$]=-0.596 shows that the best match is found for a 
$1.042+0.790$~M$_\odot$ pair at the age of 6.3~Gyr ($\log(t)=9.80$).
The primary is a red giant ($R_1=7.582$~R$_\odot$), and its parameters, 
especially the mass, are in agreement with the results from asteroseismologic 
analysis of \citet{gau14}: $M_1=1.04(12)$~M$_\odot$ and $R_1=7.14(28)$~R$_\odot$. 
The secondary is still a main-sequence ($R_2=0.770$~R$_\odot$), late G- or early 
K-type star. It's lines should be detectable on very high $S/N$ spectra.
With such a long period, this could be an interesting target for testing
the influence of rotation and activity on observed radii and temperatures
of lower-main-sequence stars. But a direct measurements of its RVs are
required.

From the difference of observed and absolute magnitudes $kmag-Kmag$, 
under the assumption of negligible extinction, we estimate the distance 
to the pair to be 807~pc. In this case the the extinction is probably not 
negligible, so the true distance is larger.

\subsubsection{KIC~10001167}
\begin{figure*}
\centering
\includegraphics[width=0.9\columnwidth]{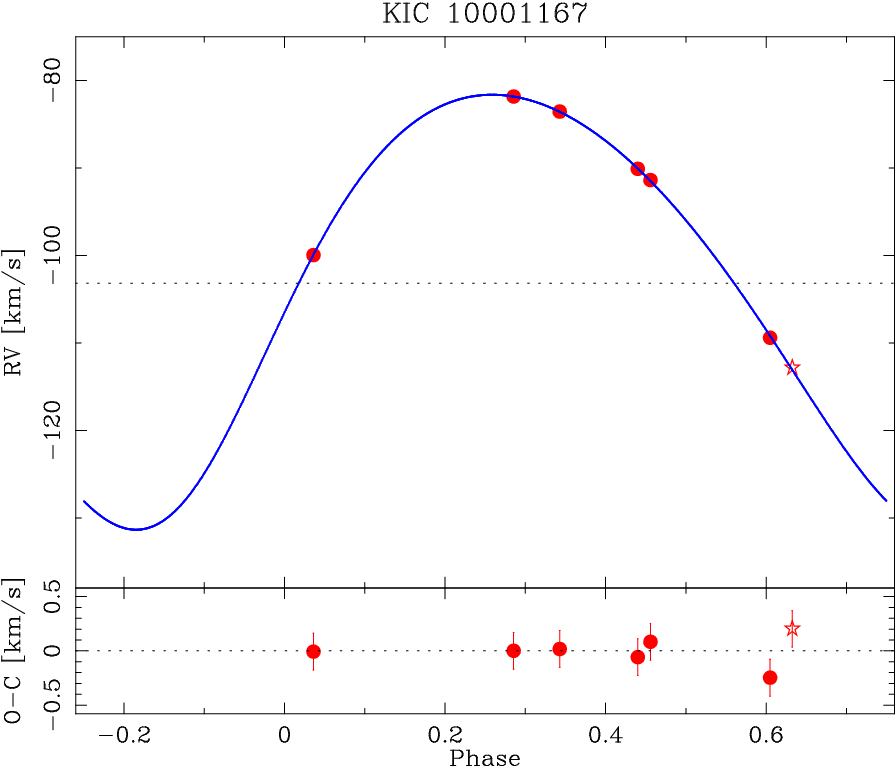}
\includegraphics[width=0.9\columnwidth]{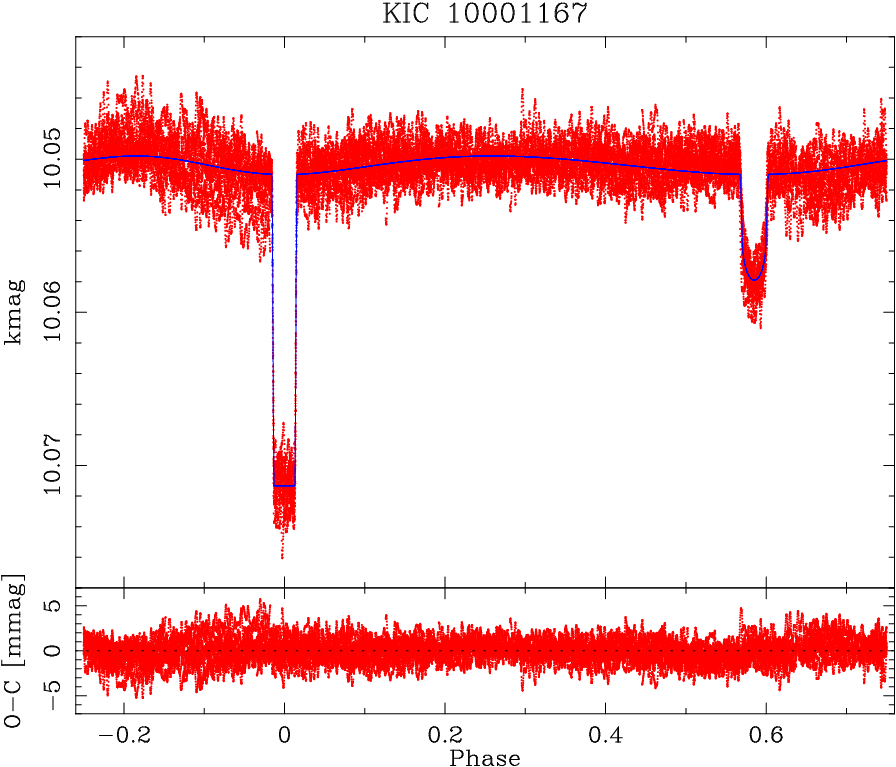}
\caption{Same as Fig.~\ref{fig_orb_0312} but 
for KIC~10001167. The increase of the velocity (left) during phase 0 
(deeper eclipse) means that the RVs are measured for the cooler 
component.
}\label{fig_orb_1000}
\end{figure*}

This system has one APOGEE and six HIDES data points, which is the lowest 
number of data points on our sample. This, and the fact that the phases where 
velocity is the lowest, are not properly covered, are the main reasons of 
high uncertainties in the RV-based parameters. The target also shows additional variability 
of the largest-scale ($rms_{LC}=1.3$~mmag), therefore the LC-based parameters
dependent on the depths of eclipses (like $i$ or $r_{1,2}$) are the least precise.
As in the previous case, the behaviour of the RV curve around the primary 
eclipse suggests that the star visible in the spectra is the cooler and
larger one. The results are shown in Table \ref{tab_par_1001}, and on Figure 
\ref{fig_orb_1000}.

The comparison with PARSEC isochrone for [$M/H$]=$-0.504$~dex showed that this
pair is likely composed of a 1.135~M$_\odot$ red giant ($R_1=14.984$~R$_\odot$),
currently during an early phase of burning helium in its core, and a 
0.993~M$_\odot$ main sequence star ($R_2=1.134$~R$_\odot$). The best-fitting
age is 4.47~Gyr ($\log(t)=9.65$). Due to large errors
in $rr, k$ and $F$, these results are however uncertain. The distance estimated 
from the difference of observed and absolute \kep magnitudes, assuming no
extinction, is 1.17~kpc. It does not agree with 705~pc estimated by \citet{gon08}.
He estimated the extinction to be 0.24~mag in $V$, which is similar to 0.29~mag,
which can be derived from the MAST entry ($E(B-V)$=0.094; assuming $R=3.1$).
Even taking 0.29~mag as the value of extinction in the \kep bandpass, 
(which should be lower than in $V$ because of the longer wavelength)
it is not enough to explain this discrepancy.

We found two other very good matches: for a $1.70+1.15$~M$_\odot$ pair at the age of 
1.6~Gyr ($\log(t)=9.2$), and for a $4.80+2.69$~M$_\odot$ pair at the age of 
0.1~Gyr ($\log(t)=8.0$). The latter solution predicted $g'-r'=0.142$~mag, which 
is much bluer than the observed 0.831, and would require very strong reddening.
The $1.70+1.15$~M$_\odot$ scenario, and the one we adopted ($1.135+0.993$~M$_\odot$),
predict $g'-r'=0.746$ and 0.752~mag, respectively. Our preferred solution is
in much better agreement with the asteroseismic parameters of the giant 
from \citet{gau14}, which are $M_1=1.13(7)$~M$_\odot$ and $R_1=13.85(32)$~R$_\odot$.
Notably, as for KIC~0843105, the mass of the primary that we found from isochrones,
matches the asteroseismologic results much better than the radius.
For the record, the $1.70+1.15$~M$_\odot$ scenario gives radii of 15.93 and
1.20~R$_\odot$. 

In any case, the  primary of KIC~10001167 is the largest star in our sample. 
The fact that it is in the core helium burning phase is interesting by itself.
KIC~10001167 seems to be a valuable target for further studies. Better understanding 
of this system would be possible if secondary's velocities were 
directly measured, and multi-band photometry obtained, especially in the total
primary eclipse. This would give direct measurement of both 
stars' masses, radii and effective temperatures. Sufficient photometry is not that 
challenging, whereas the spectroscopy would require high $S/N$ observations in the 
blue ($<$5000~\AA).

\subsubsection{KIC~10015516}

\begin{figure*}
\centering
\includegraphics[width=0.9\columnwidth]{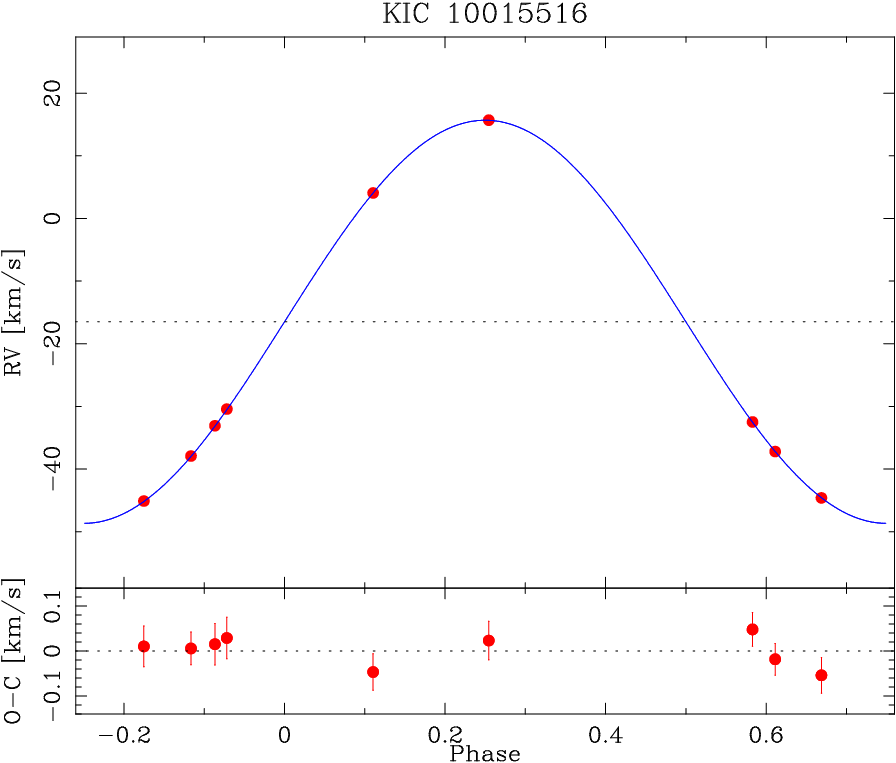}
\includegraphics[width=0.9\columnwidth]{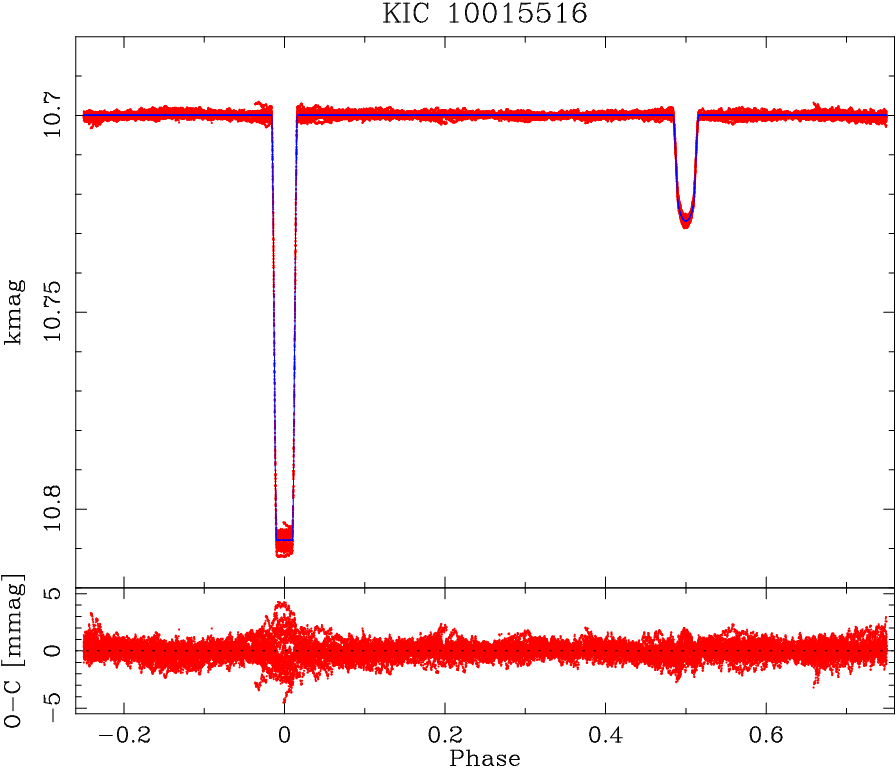}
\caption{Same as Fig.~\ref{fig_orb_0312} but 
for KIC~10015516 and our HIDES data only in the RV panel (left).
The increase of the velocity during phase 0 (deeper eclipse) 
means that the RVs are measured for the cooler component.
}\label{fig_orb_1001}
\end{figure*}

We observed this system nine times with HIDES; no APOGEE or other
publicly available spectra were found. Also in this case, the component
visible in the spectra is the cooler one, and the primary eclipse is
total. In the \jkt fit we found the orbit indifferent 
from circular, thus, to simplify our \vfit run and the RS stage
of {\sc jktebop}, we set $e$ to 0.0 and held fixed. 
Despite the largest number of degrees of freedom, we reached a very good $rms$
of the RVs of only 34 \ms.
The results are presented in Table~\ref{tab_par_1001} and in Figure~\ref{fig_orb_1001}.

This system has the largest mass function $f$ among the four SB1s with \jkt models.
It is also the only one with circular orbit, despite a relatively long period.
We thus suspected that both stars are fairly massive, the colder one being much more 
evolved than in previous cases, and the hotter, probably of type A, still residing at 
the main sequence. The best fitting PARSEC isochrone of [$M/H$]=-0.22~dex was found
for the age of 1.26~Gyr ($\log(t)=9.1$), and the preferred masses are 2.024 and
1.480~M$_\odot$ for the primary and secondary, respectively, which makes KIC~10015516
the most massive system in our sample. The estimated distance of 1.4~kpc is also
the largest from all systems described in this section. The best-fitting isochrone 
predicts the secondary to be an A8, main sequence star ($R_2=1.620$~R$_\odot$). 
The primary is found to be a red giant ($R_1=10.363$~R$_\odot$) during a late 
stage of core helium burning, just before moving to the asymptotic giant branch. 
Such configuration is a relatively unusual one for a DEB, so KIC~10015516, together
with the previous target, seems to be an valuable target for studies of very late
stages of stellar evolution. Again, better understanding of this system would 
come with direct measurements of secondary's RVs (presumably from observations in
$\lambda<5000$~\AA), and multi-band photometry.

\subsubsection{KIC~10614012 -- ``heartbeat'' SB1}
\begin{figure}
\centering
\includegraphics[width=0.9\columnwidth]{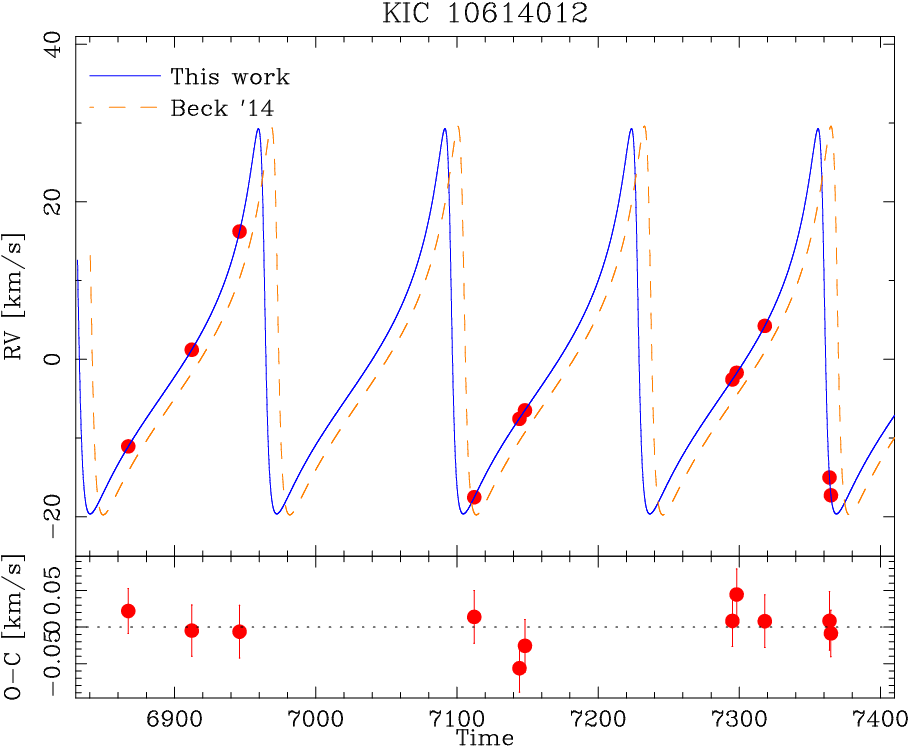}
\caption{Our HIDES RV measurements (red points) of KIC~10614012 as a function
of time, together with two different solutions: from this work (blue solid line)
and the exact solution from \citet[][orange dashed]{bec14}.
The lower panel shows residuals of our solution. 
}\label{fig_orb_1061}
\end{figure}

\begin{table}
\centering
\caption{Our orbital solution of KIC~10614012 and comparison with
the one by \citet{bec14}.}\label{tab_par_1061}
\begin{tabular}{lcc}
\hline \hline
Parameter & This work & Beck '14 \\
\hline
$P$ (d)	& 132.132(1) & 132.13(1) \\
$T_{per}$ (BJD-2450000) & 4981.38(6) & 4990.48(1) \\
$e$	& 0.709(4) & 0.71(1) \\
$\omega$ ($^\circ$) & 71.5(4) & 70.5(5) \\
$K_1$ (k\ms) & 24.46(28) & 24.68(3) \\
$\gamma$ (k\ms) & -0.67(5) &  -0.92(2) \\
$a_1 \sin(i)$ (AU) & 0.209(3) & 0.211(3)\\
$f(M_2)$ (M$_\odot$) & 0.070(2) & 0.072(3) \\ 
$rms$ (\ms) & 26 & --- \\
\hline
$M_2$  (M$_\odot$)$^a$ & 0.67(24) & 0.68(24) \\
$K_2$ (k\ms)		& 51(34) & 50(34) \\
\hline
\end{tabular}
\\$^a$ Assuming $\sin^3(i)\simeq1$ and $M_1=1.4(8)$~M$_\odot$.
\end{table}

This is an exception among our SB1s because we did not perform the \jkt 
analysis for this system. This is because \jkt is not able to model the
brightness modulation around but outside the primary eclipse, and that 
there is no secondary eclipse in the light curve. 
With HIDES we have obtained eleven spectra of this SB1, and we are able 
to compare them with the orbital solution given by \citet{bec14} on 
the basis of their 22 Hermes spectra. We found immediately that their exact 
values of orbital parameters do not reproduce our measurements. 
We first examined the \kep light curve
and found that the periods given in KEBC (132.1673~d) and by \citeauthor{bec14}. 
(132.13~d) are both not optimal, and we found a better one -- 132.132(1)~d --
similar to the one from \citeauthor{bec14}, but more precise. We performed a new 
fit to our RVs, fitting all parameters but the period (i.e.: $T_{per},e,
\omega,K_1,\gamma$), and found a satisfactory solution with $rms$ of only 
26 \ms. It is shown in Figure~\ref{fig_orb_1061}, with resulting
parameters compared in Table~\ref{tab_par_1061}.

Our new solution is quite similar to the one by \citeauthor{bec14}, but
due to the lower number of our data, and poor coverage of the phases when 
the velocity is highest and around the pericentre passage, our estimate of 
the velocity amplitude is worse. The differences in systemic velocities and 
times of pericentre may be explained by different zero-points of 
HIDES (+ATLAS~9/12) and Hermes, and the slightly different values of orbital 
periods. The marginal agreement in $\omega$ may also suggest apsidal motion,
which would also explain the difference in $T_{per}$.
We can conclude that our measurements are in general agreement with 
\citet{bec14}, but provide better constrains to the system's ephemeris.

The object itself very interesting and unusual, as it is a member
of a peculiar class of variables, called ``heartbeat stars'' (HBs), which are
objects where pulsations were tidally induced by a companion, passing closely
through the pericentre. Due to the eclipsing nature, KIC~10614012 is unusual
even among the HBs. Using the obtained values of the mass function $f(m)$, 
the estimate of the mass of the (pulsating) primary from \citet{bec14} --
1.4(8)~M$_\odot$ -- and assuming $\sin^3(i)\simeq1$ (from the existence
of eclipses), we can estimate the mass of the secondary star $M_2 = 0.67(24)$~M$_\odot$,
and the amplitude of its RV variation $K_2 = 51(34)$~k\ms. 
These estimates are also given in Tab. \ref{tab_par_1061} for each of 
the solutions. The depth of the eclipse itself is about 0.76~per~cent in
flux, which corresponds to the ratio of radii $k=\sqrt{0.0076}\simeq0.087$,
assuming negligible contribution of the secondary to the total flux.
If the primary's radius is 8.6(2)~R$_\odot$ \citep{bec14}, then the
secondary should have $\sim$0.75~R$_\odot$. It is in rough agreement with
a 0.67~M$_\odot$ main sequence star if the large uncertainty is taken into
account. Main sequence stars of this radius found in eclipsing binaries 
tend to have masses $>0.7$~M$_\odot$ \citep{sou15}.
KIC~10614012 is thus probably a pair composed of a red giant and a K dwarf.

\subsection{RV signature of solar-like oscillations in giants}\label{sec_rvosc}

\begin{figure}
\centering
\includegraphics[width=0.9\columnwidth]{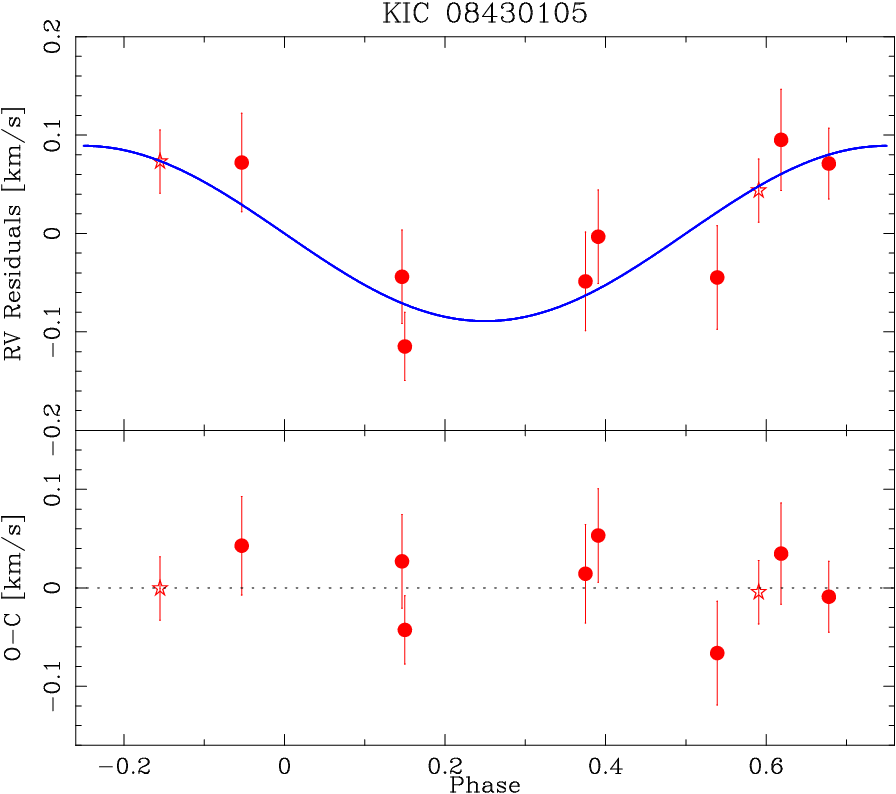}
\includegraphics[width=0.9\columnwidth]{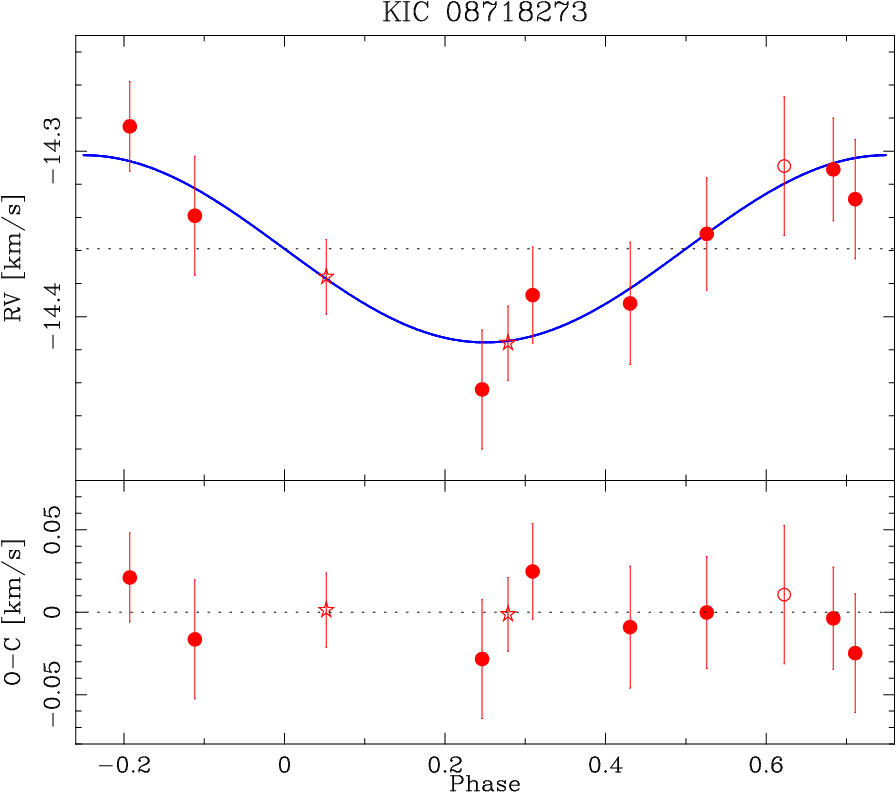}
\includegraphics[width=0.9\columnwidth]{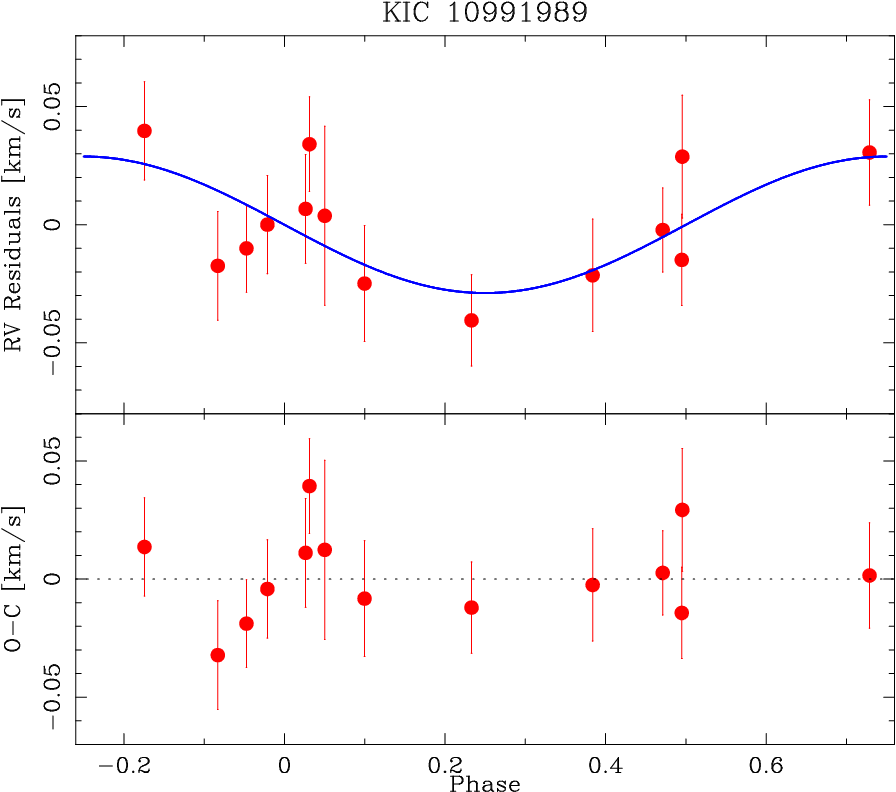}
\caption{
Our HIDES (circles) and APOGEE data (stars) for three of four oscillating red giants, 
phase folded with periods very close to the frequency of maximum amplitude.
The blue lines represent sine functions fitted 
to the data, with parameters shown in Table \ref{tab_osc}. For KIC~08718273 (middle) we show
direct RV measurements, and the open circle represents the measurement from 
the December 2014 run, shifted up by 100 \ms (see Sect. \ref{sec_rvstd}), 
matching the solution very well. 
For KIC~08430105 (top) 
and KIC~10991989 (bottom) we show residuals of the orbital fits, 
which were described earlier. 
The residuals of oscillation fits (lower panels) are shown in the same scale, for comparison.
}\label{fig_osc}
\end{figure}

\begin{table*}
\centering
\caption{Results of the search for RV signals induced by oscillations.}\label{tab_osc}
\begin{tabular}{lccccccc}
\hline \hline
	& $\nu_{max}$ & $P_{osc}$ & $P'_{osc}$ & $K_{osc}$ & $rms$ & Det. level \\
KIC No. & ($\mu$Hz) & (d) & (d) & (m s$^{-1}$) &  (m s$^{-1}$) & ($\sigma$) \\
\hline
08430105 & 66.80(1.64)& 0.1733(43) & 0.17310(2)	&  89(19)  &  71/38 & 4.60 \\
08718273 & 28.30      & 0.409      & 0.40827(3)	&  57(8)   &  68/17 & 7.12\\
10991989 & 89.90      & 0.129      & 0.12875(5) &  29(11)  &  25/19 & 2.64 \\
\hline
\end{tabular}
\\Notes: The ``$rms$'' column shows the $rms$ before and after the fit to oscillations. 
\\Uncertainty of $\nu_{max}$ was given in \citet{gau14}, but not in \citet{gau13}.
 
\end{table*}

Four of the targets described in this work are known to show solar-like 
oscillations \citep{gau13,gau14}. Such oscillations not only contribute
to the light curve of a system, but may also be affecting RVs, by introducing
an additional scatter, and possibly mimic a presence of an additional body on
a short-period orbit. We thus decided to look for their signature in our 
and APOGEE data.

For this we used the \vfit code, and assumed that the modulation is 
sinusoidal (i.e. mimicking a circular orbit). We took the frequencies of
maximum amplitude ($\nu_{max}$), and translated into periods. 
We used these values as starting 
points for the period fit. In both cases we fitted for the amplitude $K_{osc}$, 
base velocity (which corresponds to the systemic velocity $\gamma$ in KIC~08718273, 
and in the other was indistinguishable from zero), and time of the zero phase 
(which has no significant physical meaning here). 

The results are shown in Table \ref{tab_osc} and Figure~\ref{fig_osc}.
In the Table we give values of $\nu_{max}$ and its corresponding
period $P_{osc}$ (with errors, if available in the source), 
the period we found ($P'_{osc}$), and the amplitude of the RV modulation 
($K_{osc}$). We also compare the $rms$'s before and after the fit, and 
give the level of detection in $\sigma$ defined as $K_{osc}/\Delta K_{osc}$.

The signal is clearest in the case of KIC~0871273, reaching the level 
of over 7$\sigma$. The $rms$ of the 11 data points drops significantly after 
the fit, and the relative error of $K_{osc}$ is small. The second best case
-- KIC~08430105 -- also showed a significant improvement after the fit, 
but the detection level is lower (4.60$\sigma$). The last case
-- KIC~10991989 -- has he highest numbers of measurements, but the potential 
amplitude is the lowest. Detection at the level of 2.64$\sigma$ is only marginal.
For the remaining system -- KIC~10001167 -- we can only say, 
that the behaviour of the residuals, and large $rms$ of the orbital fit 
may suggest the existence of an oscillation-induced RV signal, 
but we did not find any solution that could lower the $rms$ significantly.
The number of data points is, however, still too small for a secure 
conclusion, therefore more observations are planned in the future.

We'd like to point out that the $rms$ before the fit for oscillations 
was already quite low in each case, and comparable to (for KIC~0840105 and
08718273) or better than (KIC~10991989) for the RV standards. The fit lowered
the $rms$ below 40~\ms\, (lowest $rms$ of a standard star) in each case, 
and it might be a coincidence that the periodicities we found in the data 
are very close to those expected from oscillations. Therefore we advise the 
readers to treat this result with caution.

\section{Summary}
We presented analysis of radial velocities, light curves, and eclipse 
timing variations of nine objects, found by the \kep satellite to be 
eclipsing binaries. In all cases only one set of lines has been found 
in the spectra. In our sample we identified five single-lined binaries, 
two hierarchical triples, one possible quadruple with a circumbinary
brown dwarf, and one blend. Using the 
theoretical stellar evolutionary models, we managed to estimate the absolute 
parameters of four SB1s. For the two hierarchical systems --  KIC~04758368 
and 10991989 -- we translated the ETVs into RVs and derived parameters of 
the outer orbits, together with some of the physical properties of the inner 
binaries. These objects may prove to be valuable for models of multiple stellar 
systems formation.

The target selection criteria, mainly the limits for brightness and temperature,
caused that many of our targets (6 out of 9) contain red giants, four of
which were reported to be oscillating. In our spectroscopy we were able 
to detect RV signals with periods coinciding with the oscillation frequency 
$\nu_{max}$. For two of the giants -- primary components of SB1s -- 
the isochrone-based parameters are in a good agreement with asteroseismic 
results, which suggests the correctness of our approach. 
However, notably, the parameters derived from spectroscopy, and
listed in the MAST archive, are not in agreement with ours. 
In particular, the MAST values of $\log(g)$ tend to be 0.25-0.50~dex higher 
than ours, found for four cases (see Table \ref{tab_1001_phys}). Other quantities, 
such as [$M/H$], may, therefore, also be incorrect. We thus recommend to treat the 
MAST values with bigger caution.

From the point of view of stellar astrophysics, the objects we described show
interesting properties. The components of each SB1 are clearly on very different
evolutionary stages, sometimes very rarely observed in eclipsing binaries, like the 
core helium burning phase. Unfortunately, lines of the secondaries were not visible
in our data. However, for some of them, i.e. those showing a total primary eclipse
(KIC~0843105, 10001167 and 10015516), important information can be derived from 
observations in the totality \citep{hel15a}. Further spectroscopic and photometric
follow-up during the eclipse would be very useful.

\section*{Acknowledgments}

KGH acknowledges support provided by the National Astronomical Observatory 
of Japan as Subaru Astronomical Research Fellow. M.R. acknowledges support 
provided by the National Science Center through grant 2015/16/S/ST9/00461.
This work is supported by
the Polish National Science Center grant 2011/03/N/ST9/03192, by the European 
Research Council through a Starting Grant, by the Foundation for Polish Science 
through ``Idee dla Polski'' funding scheme, and by the Polish Ministry of 
Science and Higher Education through grant W103/ERC/2011.

\appendix

\section{Radial velocities}
In this Section we list the RV measurements used in our work. We group them in
different tables, depending on their source and character. Longer tables are
available in their full extent in the on-line version of the manuscript.

In Table \ref{tab_RV_apo} we show the RV measurements extracted from the
APOGEE survey. This Table is complete. For the reference, each row contains
the Plate ID, MJD, and fibre number -- quantities useful to identify the observation
in the survey's website.

Our HIDES measurements of the nine systems described in this work
are given in Table \ref{tab_RV}. This table includes exposure 
times, final errors, and approximate $S/N$ at 6000\AA. 
Only a portion is shown here.

The Table \ref{tab_RV_etv} shows the post-ETV radial velocities, calculated for
KIC~04758368 and 10991989. Again, only a portion is shown here, and the complete
Table is available on-line.

\begin{table}
\centering
\caption{A complete set of APOGEE radial velocities used in this
study. The last column is the identifier that can be used to find 
single visit spectra in the archive.}\label{tab_RV_apo}
\scriptsize
\begin{tabular}{lcccc}
\hline \hline
BJD & $v$ & $\epsilon$ & KIC & Plate ID-MJD-Fiber \\
-2450000 & (k\ms ) & (k\ms ) & & \\
\hline
5811.612390 & -47.200 & 0.177 & 03120320 & 5213-55811-254\\
5840.592682 &  -8.638 & 0.166 & 03120320 & 5213-55840-176\\
5851.579964 & -22.274 & 0.144 & 03120320 & 5213-55851-152\\
5813.699707 & -14.087 & 0.023 & 04758368 & 5214-55813-190\\
5823.726497 & -14.439 & 0.023 & 04758368 & 5215-55823-196\\
5840.661140 & -15.046 & 0.023 & 04758368 & 5215-55840-196\\
5849.578359 & -15.419 & 0.023 & 04758368 & 5214-55849-196\\
5851.648750 & -15.584 & 0.023 & 04758368 & 5215-55851-291\\
5866.569442 & -15.902 & 0.022 & 04758368 & 5214-55866-159\\
6366.011412 &   7.517 & 0.038 & 08430105 & 6090-56365-034\\
6367.005991 &  10.042 & 0.036 & 08430105 & 6090-56366-034\\
6465.952143 & -14.376 & 0.022 & 08718237 & 7018-56465-166\\
6470.943842 & -14.416 & 0.022 & 08718237 & 7018-56470-154\\
5876.564815 &-112.834 & 0.202 & 10001157 & 5552-55876-232\\
\hline
\end{tabular}
\end{table}

\begin{table}
\centering
\caption{HIDES radial velocities. Complete Table is available
 in the on-line version of the manuscript.}\label{tab_RV}
\scriptsize
\begin{tabular}{lccccc}
\hline \hline
BJD & $v$ & $\epsilon$ & KIC & $t_{exp}$ & $S/N$ \\
-2450000 & (k\ms ) & (k\ms ) & & (s) & \\
\hline
7300.979626 & -65.177 & 0.186 & 03120320 & 1200 & 53 \\
7302.021194 & -77.543 & 0.185 & 03120320 & 1500 & 57 \\
7317.970408 &  -0.718 & 0.185 & 03120320 & 1500 & 42 \\
7490.272271 & -31.445 & 0.203 & 03120320 & 1200 & 25 \\
7491.174252 & -11.494 & 0.237 & 03120320 & 1200 & 18 \\
...\\
6866.045030 & -27.260 & 0.039 & 04758368 & 1800 & 25 \\
6912.149187 & -27.616 & 0.036 & 04758368 & 1500 & 61 \\
6914.194285 & -27.683 & 0.035 & 04758368 & 1500 & 41 \\
...\\
\hline
\end{tabular}
\end{table}

\begin{table}
\centering
\caption{Post-ETV radial velocities of centre of mass for two
hierarchical triples. Complete Table is available
 in the on-line version of the manuscript.}\label{tab_RV_etv}
\begin{tabular}{lccc}
\hline \hline
BJD & $v$ & $\epsilon$ & KIC \\
-2450000 & (k\ms ) & (k\ms ) &\\
\hline
4963.618272 &   8.51 & 2.97 & 04758368 \\
4973.452402 &   5.92 & 3.67 & 04758368 \\
4982.969772 &   7.77 & 3.34 & 04758368 \\
4991.623760 &  10.96 & 5.54 & 04758368 \\
5001.473103 &   5.26 & 2.04 & 04758368 \\
... \\
4963.617665 & -10.33 & 0.66 & 10991989 \\
4973.451707 &  -8.22 & 0.77 & 10991989 \\
4982.969009 &  -7.56 & 0.84 & 10991989 \\
4991.622949 &  -3.42 & 2.44 & 10991989 \\
5001.472255 &  -5.06 & 1.47 & 10991989 \\
... \\
\hline
\end{tabular}
\end{table}

\section{Eclipse timing variations}
In the Table \ref{tab_ETV} we show our own measurements of
eclipse timing variations for three system, as derived with the 
method described in Section \ref{sec_etv}.

\begin{table}
\centering
\caption{ The eclipse timing variations used in this work ($\tau$)
and their uncertainties $\epsilon$, calculated with the method of
\citet{koz11}. Only a portion of the table 
is shown here, and the complete one is available on-line.}\label{tab_ETV}
\begin{tabular}{lccc}
\hline \hline
BJD & $\tau$ & $\epsilon$ & KIC \\
-2450000 & (s) & (s) &\\
\hline
4958.402347 &  -606 & 68 & 04758368 \\
4968.834197 &  -494 & 61 & 04758368 \\
4978.070606 &  -532 & 65 & 04758368 \\
4987.868936 &  -476 & 64 & 04758368 \\
...\\
4955.950945 &  -0.15 & 5.88 & 05598637 \\
4960.742821 &   9.53 & 5.82 & 05598637 \\
4965.534690 &  11.16 & 7.38 & 05598637 \\
4970.326551 &   5.31 & 5.30 & 05598637 \\
...\\
4958.401792 & -141 & 31 & 10991989 \\
4968.833539 & -118 & 23 & 10991989 \\
4978.069874 & -147 & 20 & 10991989 \\
4987.868143 & -126 & 24 & 10991989 \\
...\\
\hline
\end{tabular}
\end{table}

\label{lastpage}

\end{document}